\documentclass[AMA,STIX1COL]{WileyNJD-v2}

\articletype{Article Type}%

\received{Date Month Year}
\revised{Date Month Year}
\accepted{Date Month Year}

\copyyear{2025}
\startpage{1}

\usepackage{ifpdf}
\usepackage{color, colortbl}
\usepackage{tikz}
\usepackage{array}
\usepackage{multirow}
\usepackage{listings}
\usepackage{xcolor}
\usepackage{amsmath}
\usepackage{graphicx}
\usepackage{subcaption}
\usepackage{url}
\usepackage{booktabs}
\usepackage{pifont}
\usepackage{adjustbox,siunitx}
\usepackage[table]{xcolor}  % 支持表格着色
\usepackage{booktabs}       % 美化横线
\usepackage{array}
\usepackage{booktabs}
\usepackage{float}

\lstdefinelanguage{json}{
    basicstyle=\normalfont\ttfamily,
    stringstyle=\color{darkgreen},
    literate=
     *{0}{{{\color{numb}0}}}{1}
      {:}{{{\color{punct}{:}}}}{1},
    morestring=[b]"
}
\colorlet{punct}{red!60!black}
\definecolor{background}{HTML}{EEEEEE}
\definecolor{darkgreen}{HTML}{006400}
\definecolor{delim}{RGB}{20,105,176}
\colorlet{numb}{magenta!60!black}

\begin{document}
%标题还要再改改，这里感觉不太好
% 超级平衡HBserve
\title{BanaServe: Unified KV Cache and Dynamic Module Migration for Balancing Disaggregated LLM Serving in AI Infrastructure}
\author[1,2,5]{Yiyuan He}

\author[2]{Minxian Xu}

\author[2,3]{Jingfeng Wu}

\author[1,2,5]{Jianmin Hu}

\author[5]{Chong Ma}

\author[5]{Min Shen}

\author[5]{Le Chen}

\author[4]{Chengzhong Xu}

\author[5]{Lin Qu}

\author[2]{Kejiang Ye}

\authormark{He \textsc{et al.}}
\titlemark{BanaServe: Unified KV Cache and Dynamic Module Migration for Balanced Disaggregated LLM Serving}

\address[1]{\orgdiv{Southern University of Science and Technology}, \orgaddress{\state{Shenzhen}, \country{China}}}

\address[2]{\orgdiv{Shenzhen Institutes of Advanced Technology}, \orgname{Chinese Academy of Sciences}, \orgaddress{\state{Shenzhen}, \country{China}}}

\address[3]{\orgdiv{University of Chinese Academy of Sciences}, \orgaddress{\country{China}}}

\address[4]{\orgdiv{State Key Lab of IOTSC, Faculty of Science and Technology}, \orgname{University of Macau}, \orgaddress{\state{Macau SAR}, \country{China}}}

\address[5]{\orgdiv{AIOS Team}, \orgname{Alibaba Group Inc},
\orgaddress{\state{HangZhou}, \country{China}}}

% \address[4]{\orgdiv{School of Computer Science and Engineering, South China University of Technology}, \orgaddress{\state{Guangzhou}, \country{China}}}

\corres{Minxian Xu, Shenzhen Institutes of Advanced Technology, Chinese Academy of Sciences, China \\ \email{mx.xu@siat.ac.cn}}
% 现有系统的三个限制
% 1. 静态的资源配置难以应对动态的流量变化，过高的资源预分配造成资源浪费，过低的资源预分配造成SLO违约
% 2. prefill和decode之间天然是负载不均衡的，prefill是计算密集型的，decode是显存密集型的
% 3. prefix cache aware router会根据prefix cache命中率来调度请求，会产生马太效应

\abstract[Abstract]{
Large language models (LLMs) are increasingly deployed in AI infrastructure, driving the need for high throughput, resource efficient serving systems. Disaggregated LLM serving, which separates prompt prefill from auto-regressive decode, has emerged as a promising architecture by isolating their heterogeneous compute and memory demands. However, current disaggregated systems face three key limitations: (i) static resource allocation cannot adapt to highly dynamic workloads, causing over-provisioning that wastes resources or under-provisioning that violates service level objectives (SLOs); (ii) inherent load imbalance between prefill and decode stages, where prefill is compute-bound and decode is memory-bound, causes under-utilization in one tier while the other becomes a bottleneck; and (iii) prefix cache aware routing skews load distribution, as high cache hit rate prefill nodes attract disproportionately more requests, further degrading balance and efficiency.
% Large language models (LLMs) are increasingly deployed in latency-critical applications, driving the need for high-throughput, resource-efficient serving systems. Disaggregated LLM serving, separating prompt prefill from auto-regressive decode, has emerged as a promising architecture by isolating their heterogeneous compute and memory demands. However, current disaggregated systems suffer from two fundamental limitations: (i) static resource allocation cannot adapt to highly dynamic workloads, leading to idle capacity in one stage while the other becomes a bottleneck; and (ii) prefix-caching-aware routing induces severe load skew among prefill nodes, as high hit-rate instances attract disproportionately more requests (the Matthew Effect), further degrading balance and efficiency.

To address these issues, we present BanaServe, a dynamic orchestration framework that continuously rebalances computational and memory resources across prefill and decode instances while eliminating hotspots induced by cache. BanaServe introduces layer level weight migration, attention level Key Value Cache (KV Cache) migration, and Global KV Cache Store sharing with layer wise overlapped transmission, enabling both coarse grained (layer level) and fine grained (attention level) load redistribution with minimal latency overhead. These mechanisms allow routers to perform purely load aware scheduling, unconstrained by cache placement.

We implement BanaServe on top of state-of-the-art LLM serving stacks (e.g. vLLM and DistServe) and evaluate it on diverse workloads including long-context inference, bursty query arrivals, and mixed prompt–generation patterns. Compared to vLLM, BanaServe achieves \textbf{1.2×–3.9×} higher throughput with \textbf{3.9\%–78.4\%} lower total processing time, and outperforms DistServe by \textbf{1.1×–2.8×} in throughput with \textbf{1.4\%–70.1\%} latency reduction. These results demonstrate that BanaServe delivers substantial gains under dynamic, real-world LLM serving conditions.
}
\keywords{Large Language Models, Cloud Computing, Disaggregated LLM Serving, Resource Management, AI Infrastructure}

\maketitle

% 1.介绍LLM Serving 
% 2.介绍LLM的forward过程，解释什么是layers和attention计算以及KVCache和后面使用的各种指标：TTFT，TPOT，e2e latency
% 3.介绍常见的加速手段：
    %3.1 PD分离，介绍PD分离相关的工作：DistServe，loongserve，以及serverlessLLM（画个pd分离的架构图）
    %3.2 prefix cahcing，解释什么是prefix caching，怎么加速的，以及目前的prefix cache aware router => SGlang 
% 4.目前主流系统的问题：（这两个可以画两个图来说明问题）
    %4.1 问题1: 静态的配置/冷启动问题难以应对brust，并且产生资源碎片，并且存储资源利用不均衡，prefill使用cpu memeory，decode使用HBM，prefill中的HBM仅仅作为缓存，decode是memory bound，prefill 是compute bound
    %4.2 问题2: prefix cache aware router 会产生马太效应，prefill实例之间会产生负载的不均衡
% 5.介绍banaserve，我们在decode 和 prefill 阶段各自定制了两个操作迁移weight和迁移KV去balance 计算和显存
% 6.我们的贡献：
% - 我们发现了静态的资源配置难以适应动态的请求，以及prefix cache aware router 会导致prefill节点之间负载不均衡的问题
% - 我们提出了banaserve解决了上面的两个问题，banaserve通过权重迁移 和KV Cache ，以及attention的卸载，去动态平衡系统的资源开销适应动态的请求（问题1），并且采用统一的KV Cache 池，采用layer wise 的传输方式去overlap通信开销，从而使所有的prefill共享KV Cache，router只需要关注负载，无需考虑prefix cache。
% - 我们实现了banaserve并且和distserve，vllm （这里可能再增加一些sota）性能对比提升了多少
\section{Introduction}\label{sec:introduction}

% 1.介绍LLM Serving 
The rapid advancement of Large Language Models (LLMs) such as GPT-4~\cite{achiam2023gpt}, LLaMA~\cite{touvron2023llama}, and Claude~\cite{claude} has revolutionized applications in conversational AI, code generation, knowledge retrieval, and content creation \cite{monteiro2025nocodegpt,nguyen2025generative}. With models containing billions of parameters \cite{zeng2025subkv}, serving LLMs efficiently at scale presents significant challenges: maintaining low latency for interactive applications, achieving high throughput for cost-effectiveness, and maximizing hardware utilization under dynamic workloads~\cite{wen2025statuscale}. These challenges are particularly acute as LLMs transition from research prototypes to production systems serving millions of users.

% 2.介绍LLM的forward过程，解释什么是layers和attention计算以及KVCache和后面使用的各种指标：TTFT，TPOT，e2e latency

Most state-of-the-art LLMs adopt the decoder-only Transformer architecture~\cite{vaswani2017attention}, whose forward computation during inference can be conceptually divided into two sequential stages: the prompt prefill stage and the auto-regressive decoding stage~\cite{zhou2024survey}. In the prompt prefill stage, the model processes the entire input sequence in one pass, allowing all Transformer layers to operate over all tokens in the prompt simultaneously. In the auto-regressive decoding stage, the model generates output tokens one at a time, with each new token conditioned on all previously generated tokens. This computational pattern is dictated by the self-attention mechanism, which requires every token to attend to all tokens before it. To mitigate redundant computation in the decoding stage, modern serving systems employ the KV Cache \cite{zhang2023h2o,ge2023model} to store intermediate attention states of past tokens, enabling efficient reuse across steps. These two stages lead naturally to several key performance metrics for evaluating LLM serving. The Time to First Token (TTFT) measures the latency from request arrival to the generation of the first token. TTFT is mainly determined by the prompt prefill stage. The Time Per Output Token (TPOT)~\cite{zhong2024distserve} reflects the average time required to generate each subsequent token during decoding. The end-to-end latency captures the total time from request submission to the completion of the entire output sequence.

% 为了优化llm serving，目前主流的加速方式为pd分离和prefix caching。
% 在这里介绍这两个的基本原理和相关的paper和系统
% 分析这两个引入的问题：负载不均衡。 
% pd分离的负载不均衡：首先pd分离的系统天然是负载不均衡的，这是因为prefill是计算密集型的，decode是计算密集型的。其次对于动态的请求变化，pd分离会产生一些资源碎片难以被利用（这里可以解释一下为什么：适应动态的请求，需要动态的pd配比，但是pd配比的粒度很大而且速度慢在有限的资源中，难以有效的进行动态调整pd配比，去适应动态的请求变化）
% prefix caching aware router（这里简单介绍一下这个router：把请求调度到prefix caching命中率最高的prefill实例中） 会产生马太效应效应，进而导致prefill实例间的负载不均衡。

To accelerate LLM serving, recent research has proposed a number of system level optimizations, among which two representative techniques are prefill/decoding (PD) disaggregation and prefix caching. PD disaggregation, as adopted by systems such as DistServe~\cite{zhong2024distserve} and Splitwise~\cite{patel2024splitwise}, assigns the prompt prefill stage and the auto-regressive decoding stage to different sets of GPUs. This design eliminates interference between prefill and decode by preventing the two stages, whose resource demands differ significantly, from competing on the same device.

On the other hand, prefix caching, as implemented in systems like SGLang~\cite{sglang} and vLLM~\cite{vllm}, stores the KV Cache corresponding to common prompt prefixes and reuses it across requests. Combined with a prefix cache aware router that dispatches requests to prefill instances with the highest cache hit rates, this approach can substantially reduce redundant computation in the prefill stage and improve throughput in high load scenarios.

\begin{figure}[htbp]
    \centering
    % Prefix Cache Router
    \begin{subfigure}[b]{0.5\textwidth}
        \centering
        \includegraphics[width=\textwidth]{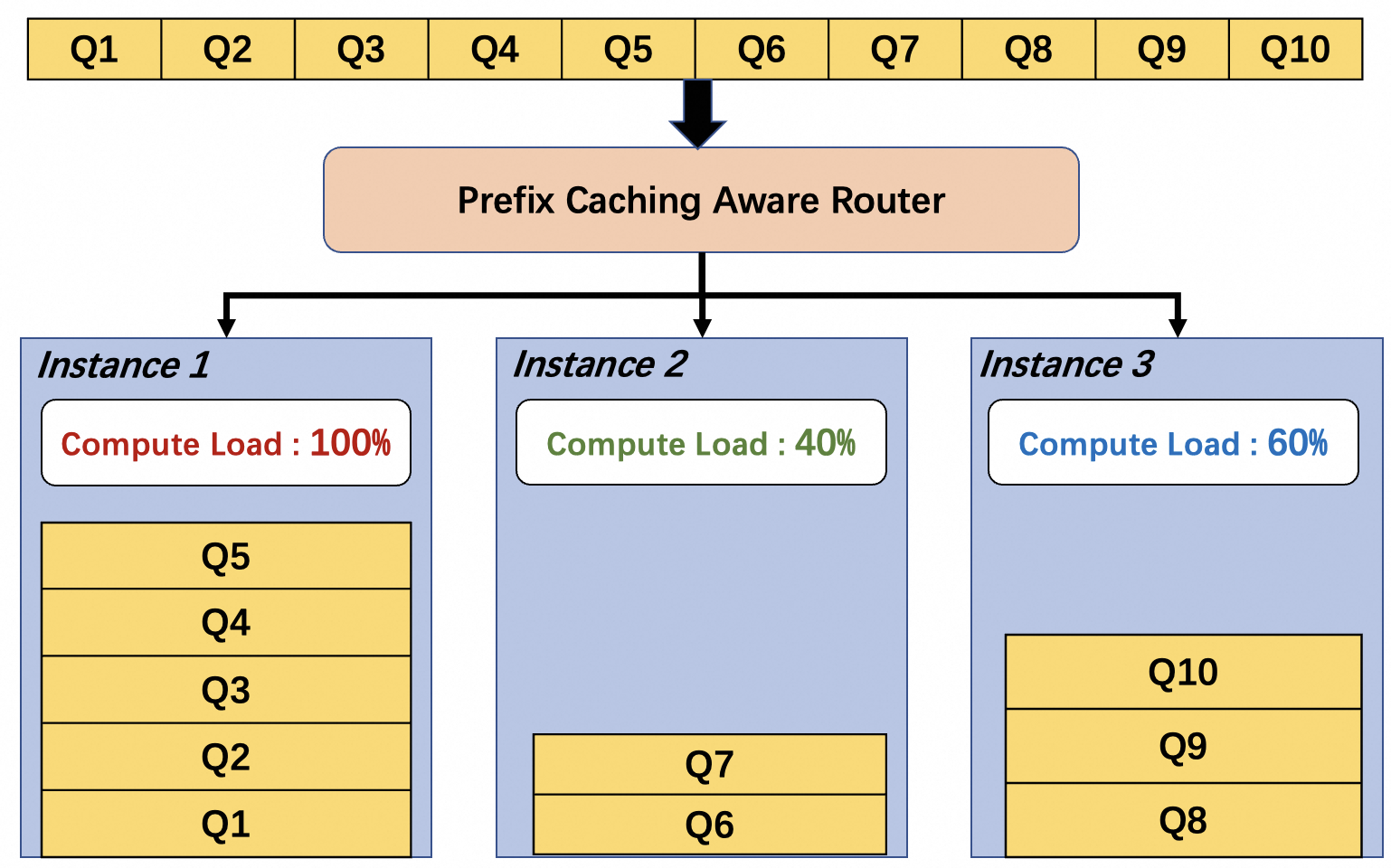}
        % \caption{Load imbalance caused by prefix cache aware router. Requests with popular cached prefixes are directed to a small set of prefill instances (e.g., Instance~1 at 100\% compute load), while others remain underutilized. This cache-affinity-driven policy results in a ``Matthew effect'' that exacerbates tail latency and reduces throughput.}
        \caption{Load imbalance caused by a prefix cache aware router. 
        Requests associated with popular cached prefixes are preferentially directed to high-hit-rate instances, 
        such as Instance~1, which operates at full compute load (100\%) and stores prefixes Q1--Q5. 
        Instance~2 remains at lower utilization (40\%) while holding redundant entries Q6--Q7 that appear in other caches. 
        Instance~3 sustains 60\% load but recomputes uncached prefixes Q8--Q10. 
        This allocation pattern leads to persistent skew, redundant storage, and repeated computation.}
        \label{fig:prefix_cache_router}
    \end{subfigure}
    \hfill
    % PD Imbalance
    \begin{subfigure}[b]{0.48\textwidth}
        \centering
        \includegraphics[width=\textwidth]{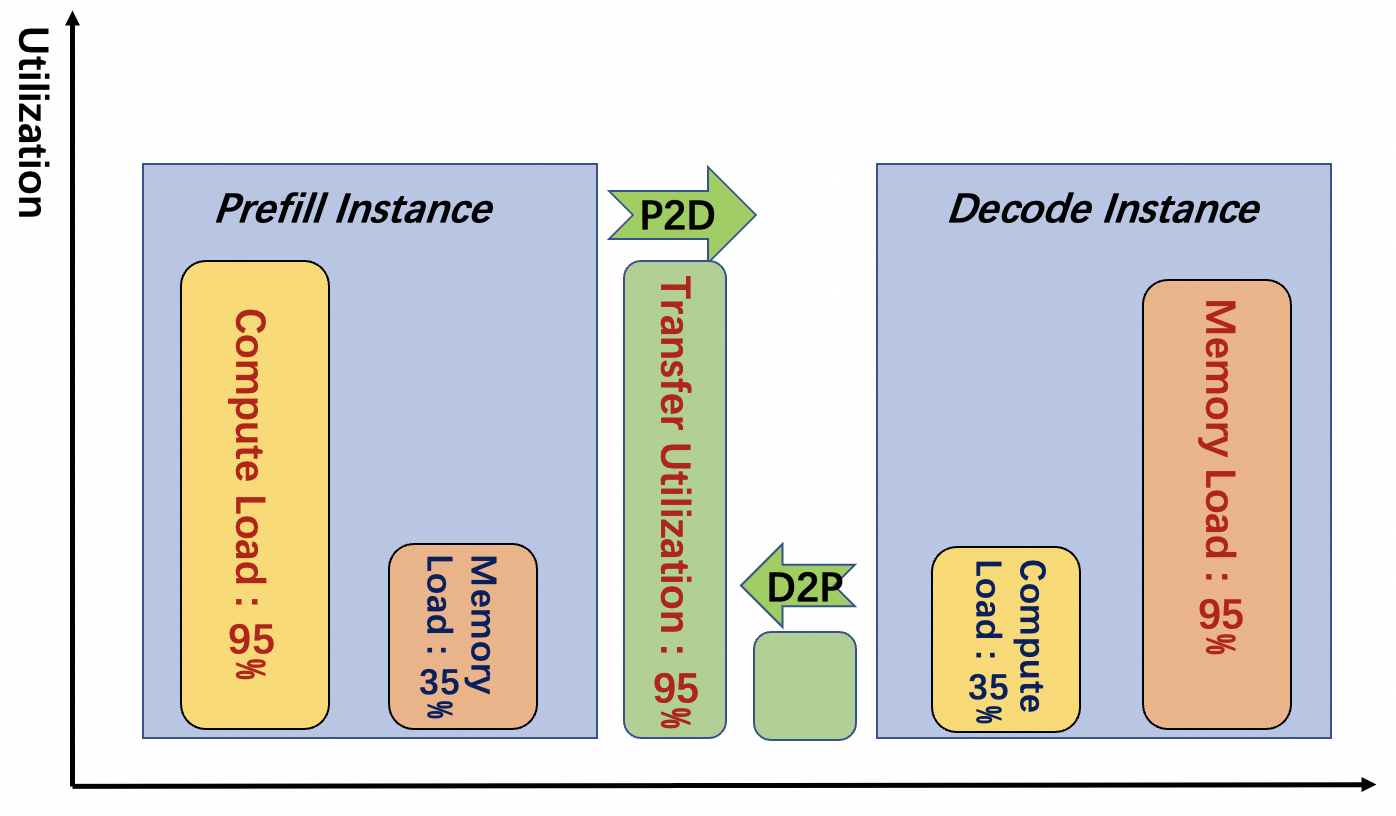}
        % \caption{Resource utilization asymmetry in PD disaggregated architectures. Prefill instances sustain high compute utilization ($\approx$95\%) but low memory utilization ($\approx$35\%), while decode instances exhibit the opposite pattern. Prefill$\rightarrow$Decode bandwidth dominates, leaving other communication underutilized.}
        \caption{Resource utilization asymmetry in PD disaggregated architectures measured using the DistServe framework with a LLaMA-13B model on the Alpaca dataset. 
        Prefill instances processing long prompts sustain high compute utilization (approximately 95\%) but low memory usage (approximately 35\%), 
        whereas decode instances exhibit the opposite pattern with lower compute utilization (approximately 35\%) and high memory usage. 
        Communication is dominated by one-way KV Cache transfers from prefill to decode, leaving other channels underutilized.}
        \label{fig:pd_imbalance}
    \end{subfigure}

    % \caption{Illustrations of two fundamental limitations in current LLM serving architectures: (a) load skew caused by cache-aware routing; (b) compute, memory and bandwidth asymmetry inherent to PD disaggregation.}
    \caption{Two empirically validated limitations in current LLM serving architectures.  
    (a) Prefix-cache-aware routing leads to persistent load skew, redundant cache storage and repeated computation, as high-hit-rate instances attract more requests while others remain underutilized.  
    (b) PD disaggregation exhibits an inherent imbalance across compute, memory and bandwidth resources, resulting in under-utilization of one dimension while another becomes a bottleneck.}
    \label{fig:limitations}
\end{figure}

However, both strategies exhibit fundamental limitations under realistic, dynamic workloads:
\begin{enumerate}

\item \textbf{Static configurations lead to resource under-utilization.} 
Modern LLM serving systems frequently exhibit sub-optimal resource usage under static allocation policies, driven by two primary factors. First, at low request rates (RPS $\leq$ 10), substantial portions of computational and memory capacity remain idle. 
As shown in Fig.~\ref{fig:bg1}, empirical measurements reveal that Hugging Face Transformers (HFT)~\cite{transformers} and vLLM~\cite{vllm} leave approximately 20\%--40\% of GPU resources unused under these conditions. Second, a fundamental mismatch between fixed resource allocations and the dynamic workload characteristics of LLM serving further exacerbates inefficiency. 
In traditional designs, meeting storage requirements often constrains available compute capacity, and vice versa, preventing balanced utilization across resource types. This persistent allocation–demand disparity motivates our design of a dynamic resource allocation mechanism to enhance the cost-effectiveness of deployment.

%
%像sglang这种使用prefix cache aware router的系统很容易出现这种问题，并解释原因：prefix cache aware router会同时考虑prefix cache的命中率和负载情况，会导致命中越高的实例越容易被调度，命中越低的实例越容易出现重计算现象。并且各个实例之间存储了很多冗余的数据
\item \textbf{Prefix cache aware router induces severe load imbalance.} 
As shown in Figure~\ref{fig:prefix_cache_router}, a representative scenario illustrates how a prefix cache aware router allocates incoming requests across three serving instances based on both cache hit rate and current load. 
Although this policy seeks to balance computation with cache locality, it systematically favors instances with high prefix cache hit rates, resulting in persistent load skew and inefficient resource usage.

In this example, Instance~1 achieves the highest cache hit rate and receives the majority of incoming queries. 
It operates at full compute capacity (\(100\%\)) while holding a large set of cached prefixes (Q1–Q5), reinforcing its scheduling priority. 
Instance~2 receives fewer queries and maintains low compute utilization (\(40\%\)), its cached prefixes (Q6–Q7) also reside in other instances, creating redundant KV Cache storage across the cluster. 
Instance~3 sustains moderate compute utilization (\(60\%\)) but primarily serves uncached prefixes (Q8–Q10), which must be recomputed due to their absence in high-priority caches, thereby incurring additional latency and forfeiting cache benefits.

This allocation pattern highlights a positive-feedback dynamic: 
instances with high hit rates attract more requests and expand their cache content, while lower-hit-rate instances process fewer queries yet store duplicate data or perform repeated computations. 
Over time, this imbalance increases queuing delays and degrades cache efficiency on overloaded instances, while under-utilized devices fail to contribute fully to overall system throughput.
\begin{figure}
    \centering
    \includegraphics[width=0.55\linewidth]{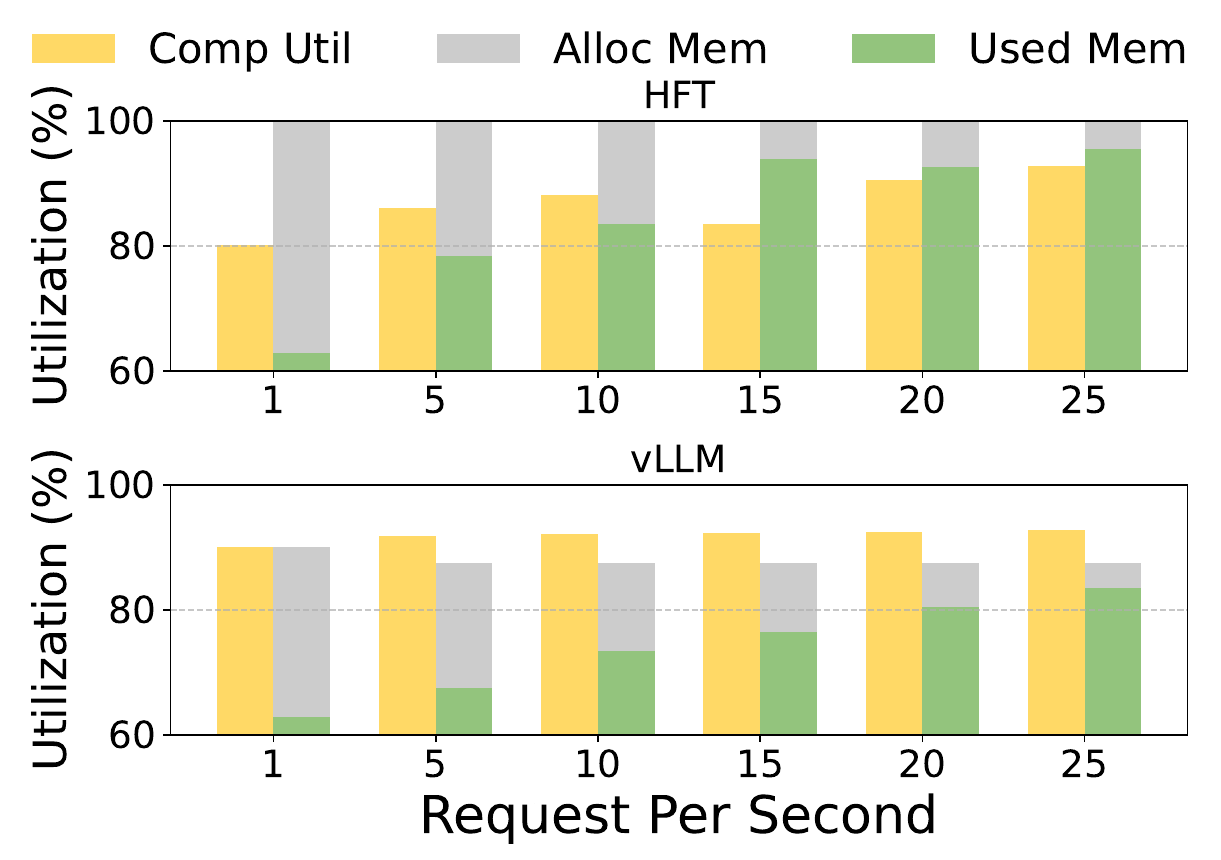}
    \caption{GPU resource utilization comparison between HFT and vLLM serving frameworks across different request rates, conducted with a single LLaMA-13B instance deployed on an A100 GPU, with each test repeated five times.}
    \label{fig:bg1}
    \vspace{-1em}
\end{figure}

% Figure~\ref{fig:prefix_cache_router} illustrates how prefix cache aware routers produce a "Matthew effect", meaning that the already advantaged gain further advantages while the disadvantaged fall further behind. In this context, instances with popular cached prefixes attract disproportionately more requests (Instance~1 at 100\% load), while others remain underutilized (Instance~2 at 40\%). This imbalance emerges because routing decisions prioritize cache locality over load distribution. Over time, the imbalance intensifies: heavily loaded instances suffer increased queuing delays and degraded cache performance, yet the router persists in directing traffic to them due to cache affinity. As a result, tail latency grows and overall system throughput declines, with some GPUs overloaded while others have spare capacity.

%pd分离时不同，prefill和decode是天然不均衡的，如Figure~\ref{fig:pd_imbalance}，prefill实例往往需要处理长文本的prefill，对算力要求更高，decode实例只需要进行自回归，并且使用KV Cache减少计算量，这就导致在pd分离的系统中，prefill实例的计算负载更高，往往能达到95%以上，但是显存负载比较低，一般在35%左右。与之相反的是decode实例，decode实例的计算负载往往在35%左右，显存负载会更高。此外在带宽利用方面，prefill实例需要将preifll阶段产生的KV Cache和next token传给decode实例，因此只充分利用了prefill到decode实例的单向通信。这种天然的不均衡也导致了pd分离架构难以有效的利用资源。
\item \textbf{PD disaggregation introduces intrinsic load imbalance.} As shown in Figure~\ref{fig:pd_imbalance}, this imbalance is derived from empirical measurements conducted in our experimental environment. 
Specifically, we deployed the DistServe~\cite{zhong2024distserve} with a LLaMA-13B~\cite{touvron2023llama} model, using the publicly available Alpaca~\cite{alpaca} dataset to generate inference workloads. 
We instrumented both the prefill and decode instances to record compute and memory utilization during execution.

The results show that the prefill stage typically processes long prompt sequences, making it highly compute intensive (often sustaining $>$95\% compute utilization) but relatively light on memory utilization (around 35\%). 
In contrast, the decode stage executes iterative auto-regressive generation and leverages the KV Cache generated during prefill to reduce computation, leading to lower compute utilization (around 35\%) but significantly higher memory utilization. 
This asymmetry is further reflected in inter-stage communication: prefill instances must transmit KV Cache data and next-token outputs to decode instances, resulting in predominantly one-way bandwidth usage from prefill to decode. 
Such empirically observed load imbalance in PD disaggregated architectures limits effective resource utilization across devices, leaving one resource dimension underused while the other becomes a performance bottleneck.

% As shown in Figure~\ref{fig:pd_imbalance}, the prefill stage typically processes long prompt sequences, making it highly compute intensive (often sustaining $>$95\% compute utilization) but relatively light on memory utilization (around 35\%). In contrast, the decode stage performs iterative auto-regressive generation and leverages the KV Cache generated during prefill to reduce computation, leading to lower compute utilization (around 35\%) but significantly higher memory utilization. This asymmetry is further reflected in inter-stage communication: prefill instances must transmit KV Cache data and next token outputs to decode instances, resulting in predominantly one way bandwidth usage from prefill to decode. Such inherent load imbalance in PD disaggregated architectures prevents effective resource utilization across devices, leaving one resource dimension underused while the other becomes a performance bottleneck.

\end{enumerate}

% These limitations stem from a fundamental tension in current architectures: the coupling of resource allocation decisions with caching policies. Static PD disaggregation cannot adapt because moving computation requires moving state; cache-aware routing cannot balance load because it must preserve cache locality.
These limitations share a common root cause: the tight coupling between resource allocation and state (cache) placement in current architectures. In PD disaggregation, computational and memory resources are statically partitioned across prefill and decode instances, but the state required to execute a stage, such as weights for computation or KV Cache for memory, cannot be easily relocated without incurring significant overhead. As a result, adjusting resources dynamically in response to workload shifts becomes impractical, because moving computation inevitably entails moving large volumes of associated state. Similarly, cache aware routing is constrained by locality: to maximize cache hit rates, it must direct requests to instances holding the relevant prefixes, even if those instances are overloaded. This perpetuates load imbalance, as high hit rate nodes continuously attract more traffic while low hit rate nodes remain underutilized. Together, these constraints prevent current systems from simultaneously achieving high resource utilization and balanced load, particularly under dynamic and unpredictable workloads.

To address these challenges, we present \textbf{BanaServe}, a dynamic orchestration framework that decouples resource allocation from state management through three key innovations. First, BanaServe introduces \emph{fine-grained resource migration} mechanisms that include layer-level weight migration, attention-level KV Cache migration, and selective computation offloading, enabling rapid rebalancing between prefill and decode instances without service disruption. Second, BanaServe employs a \emph{Global KV Cache Store} accessible to all prefill instances, eliminating cache locality constraints on routing decisions. Third, BanaServe implements \emph{layer-wise overlapped transmission} that hides communication latency by pipelining KV Cache transfers with layer computation, making cache sharing practical at scale.

These mechanisms work synergistically to enable truly adaptive serving: the orchestrator continuously monitors system load and dynamically redistributes resources at fine granularity, while the router makes purely load-aware scheduling decisions unconstrained by cache placement. This design philosophy of separating mechanism (resource allocation) from policy (routing) allows BanaServe to achieve both high efficiency and robust performance under diverse workload conditions.

Our main \textbf{contributions} are summarized as follows:
\begin{enumerate}
\item  We reveal two intrinsic limitations of current LLM serving systems: (i) static resource configurations fail to adapt to non-stationary workloads; (ii) inherent load imbalance between prefill and decode stages, where prefill is compute bound and decode is memory bound, causing under-utilization in one tier while the other becomes a bottleneck; and (iii) prefix cache aware routing inherently skews load distribution.
\item We design and develop \textbf{BanaServe}, a dynamic orchestration system for LLM serving that decouples resource allocation from state management. The system architecture enables adaptive rebalancing between prefill and decode instances under highly dynamic workloads.
\item Within BanaServe, we present several key mechanisms: dynamic weight migration, KV Cache offloading, and attention computation offloading for fine-grained resource balancing, and a Global KV Cache Store with layer-wise transmission to support cache sharing and communication–computation overlap.
% \item  We propose BanaServe, which introduces dynamic weight migration, KV Cache offloading, and attention computation offloading to achieve adaptive resource balancing, as well as a Global KV Cache Store with layer-wise transmission to enable cache sharing and communication–computation overlap.
\item We implement BanaServe and evaluate it with 13B-parameter models on production workload traces. Compared to vLLM~\cite{vllm}, BanaServe achieves up to 3.9× higher throughput with 78.4\% lower latency, and compared to DistServe~\cite{zhong2024distserve}, it delivers 2.8× throughput improvement with 70.1\% latency reduction, while maintaining robust performance across both short-context (Alpaca~\cite{alpaca}) and long-context (LongBench~\cite{longbench}) scenarios.
\end{enumerate}

\section{Background}\label{sec 2}

To understand the challenges addressed by BanaServe, we first outline the computational characteristics of prefill and decode in LLM inference, then evaluate limitations of both co-located and disaggregated deployment architectures.

% pd 的特征
\subsection{Characteristics of Prefill and Decode Operations}

Prefill and decode phases in LLM inference have fundamentally different execution characteristics and resource requirements.  During the \emph{prefill} phase, the model processes the entire input prompt in parallel across the sequence dimension \cite{dao2022flashattention,shah2024flashattention}, enabling high degrees of parallelism in matrix multiplications and attention computation. This stage is thus \emph{compute-bound}, saturating GPU cores and delivering predictable runtimes primarily determined by input length and model size.

In contrast, the \emph{decode} phase generates new tokens sequentially in an autoregressive fashion, making each step dependent on the previous output. This inherently limits parallelism, shifting the bottleneck from compute throughput to \emph{memory-bound}. The decode stage involves frequent accesses and updates to the KV Cache, resulting in irregular and memory-intensive access patterns. Furthermore, output lengths vary across requests, introducing uncertainty and making decode workloads harder to predict and balance.

% pd 不分离会有哪些问题：TPOT会被prefill拖慢，导致TPOT和e2elatecny无法满足SLO
\subsection{Inefficiencies of Co-located Prefill and Decode}

The co-location of prefill and decode on the same GPU instances causes fundamental inefficiencies.  
First, the mismatch in computational characteristics leads to resource contention: compute-bound prefill interferes with memory-bound decode, inflating latency for both. When prefill saturates GPU compute resources, decode requests are delayed, while decode-heavy intervals leave computational throughput underutilized.

Second, co-location architectures~\cite{mukherjee2023orca} often maintain a one-to-one mapping between prefill and decode threads for pipeline simplicity. However, because \textsc{TTFT} is often orders of magnitude larger than \textsc{TPOT}, the slower prefill phase throttles overall throughput and prevents decode from reaching full utilization.

Finally, resource utilization patterns exhibit substantial volatility throughout the inference lifecycle. During the prefill phase, both memory consumption and compute utilization peak due to attention computation and KV Cache initialization. However, once the system transitions to decode, memory usage patterns shift dramatically and compute requirements decrease significantly. These opposed patterns make it difficult to provide a fixed resource allocation that avoids both under-utilization and bottlenecks.

% The co-location of prefill and decode operations on the same instances introduces several critical inefficiencies. First, the fundamental mismatch in computational characteristics leads to resource contention and suboptimal utilization. When prefill operations dominate GPU compute resources, decode operations experience increased latency due to scheduling delays and resource competition. Conversely, during decode-heavy periods, valuable compute resources remain underutilized.

% The architectural requirement of maintaining one-to-one correspondence between prefill and decode instances in co-located systems creates throughput saturation issues. Since TTFT is typically orders of magnitude larger than TPOT, the system's overall throughput becomes limited by the slower prefill phase, preventing the decode phase from reaching its full potential. This imbalance results in significant underutilization of decode capacity.

% Resource utilization patterns exhibit substantial volatility throughout the inference lifecycle. During the prefill phase, both memory consumption and compute utilization peak due to attention computation and KV Cache initialization. However, once the system transitions to decode, memory usage patterns shift dramatically, and compute requirements decrease significantly. This fluctuation makes it challenging to provision resources efficiently and maintain consistent performance.

% pd 分离后导致资源利用不均衡
% 算力使用不均衡：prefill阶段的MFU远大于decode阶段的MFU，prefill的算力容易出现瓶颈
% 显存利用不均衡：prefill阶段的显存限制在整个的集群的CPU host memory，decode 的显存限制在GPU HBM，往往decode的显存容易出现瓶颈
% 这里的图参考blitzScale

\subsection{Imbalanced Utilization in PD Disaggregation}
While PD disaggregation addresses the interference issues of co-located serving, it introduces new challenges related to resource imbalance between prefill and decode instances. Previous work, exemplified by Taming~\cite{taming} and Blitzscale~\cite{blitzscale}, has documented substantial differences in resource utilization across dedicated prefill and decode clusters. In particular, GPU utilization monitoring in Sarathi-Serve~\cite{taming} shows that prefill instances often operate at high capacity during peak periods but experience substantial idle time during low-traffic intervals. Conversely, decode instances may become bottlenecks when handling long-running generation tasks, even as prefill instances remain underutilized.

Our own measurements (Figure~\ref{fig:pd_imbalance}) on representative LLM workloads corroborate these findings, revealing similar asymmetries in both compute and memory utilization. Prefill instances require substantial memory for processing large input sequences and initializing KV Cache, but this memory becomes available once requests are transferred to decode instances. Decode instances, meanwhile, accumulate KV Cache memory over time as generation proceeds, creating asymmetric memory pressure across the system. This imbalance manifests in scenarios where prefill instances may be memory-constrained while decode instances have abundant free memory, or vice versa.
% TODO
% 动态的资源变化导致pd不同的算力和显存的需求，需要动态调整pd的资源配置。
% 这里可以画个图，表示在使用brustGPT的benchmark 数据集里，随着请求的QPS的变化，prefill需要的资源（算力，显存）的变化，以及decode 需要的资源（显存、算力）的变化
\subsection{Limitations of Static PD Disaggregated Configuration under Dynamic Workloads}
Current PD disaggregation architectures rely on predetermined, static configurations that fail to adapt to the dynamic nature of real-world workloads. During periods of low request rates, the system suffers from resource over-provisioning, with dedicated prefill and decode instances remaining underutilized. In contrast, during high-traffic periods, the static allocation becomes a throughput bottleneck, unable to scale beyond the predetermined capacity limits.

Sudden traffic spikes present particularly challenging scenarios for static configurations. Bursty request patterns can overwhelm individual prefill or decode nodes, potentially leading to system failures or severe performance degradation. The lack of elasticity in static systems means that temporary traffic surges cannot be absorbed by redistributing load across available resources, resulting in poor fault tolerance and service reliability.

The operational overhead of adjusting prefill and decode instance configurations is prohibitively expensive in static systems. Scaling operations typically require service restarts, model reloading, and careful coordination to maintain consistency. This rigidity prevents systems from responding quickly to changing workload patterns and limits the ability to optimize resource allocation based on real-time performance metrics.

\section{Related Work}\label{sec 3}

In this section, we review prior work on phase-disaggregated system architectures, with a focus on their application to LLM serving systems. We also survey complementary research directions, including request scheduling strategies and dynamic load balancing techniques, which collectively address key challenges in distributed inference.

% 增加新的llm serving ： VLLM，MII，TRT-LLM，dynamom，sglang
\subsection{LLM Serving}
Recent advances in LLMs serving aim to reduce inference latency and improve hardware utilization for models with billions of parameters, which must process long-context prompts and perform autoregressive decoding under strict SLOs. In this setting, the dual-phase execution pipeline, comprising the compute bound prefill phase and the memory bound decode phase, creates challenges for sustaining high throughput, minimizing tail latency, and efficiently managing KV Cache memory.

Several representative systems have been proposed to improve inference efficiency within monolithic serving architectures. vLLM~\cite{vllm} introduces PagedAttention, a block-level KV Cache management mechanism that reduces memory fragmentation and enables efficient batched decoding \cite{mukherjee2023orca} for dynamic, variable-length sequences. SGLang~\cite{sglang} focuses on optimizing prompt processing through prefix caching and a cache-aware scheduling policy, which routes requests to maximize KV Cache hit rates and thereby reduce redundant computation in the prefill phase. DeepSpeed-FastGen~\cite{holmes2024deepspeed} builds on DeepSpeed-Inference by incorporating optimizations for high-throughput, low-latency text generation, including enhanced kernel fusion, improved batching strategies, and adaptive communication scheduling to maximize performance across diverse workload patterns. TensorRT-LLM~\cite{tensorrtllm} incorporates graph-level compilation, fused attention kernels, and quantization-aware execution to deliver low-latency, high-throughput Transformer inference on NVIDIA GPUs. Extending this stack to distributed environments, NVIDIA Dynamo~\cite{nvidiadynamo} serves as a multi-node orchestration layer atop TensorRT-LLM, providing elastic scaling, heterogeneous resource scheduling, and coordinated KV management for large-scale enterprise deployments.

While these systems achieve substantial performance gains through kernel optimization, memory management, and scheduling mechanisms, they generally co-locate the prefill and decode phases on the same set of devices. Such tight coupling can exacerbate interference between the prefill and decode phases, where the divergent computational characteristics of each phase compete for GPU compute cycles and memory bandwidth. Moreover, static resource allocation in these monolithic designs hampers adaptability to dynamic and bursty workload patterns, leading to persistent under-utilization in one phase while the other becomes a throughput bottleneck. These observations motivate research into phase-disaggregated serving architectures, which physically or logically separate the prefill and decode stages to enable optimization specific to each phase. This direction is reviewed in the following subsection.

% While these systems achieve substantial performance gains through kernel optimization, memory management, and scheduling mechanisms, they generally co-locate the prefill and decode phases on the same set of devices. Such tight coupling can exacerbate PD interference, where the divergent computational characteristics of each phase compete for GPU compute cycles and memory bandwidth. Moreover, static resource allocation in these monolithic designs hampers adaptability to dynamic and bursty workload patterns, leading to persistent underutilization in one phase while the other becomes a throughput bottleneck. These observations motivate research into phase-disaggregated serving architectures, which physically or logically separate the prefill and decode stages to enable phase-specific optimization—a direction reviewed in the following subsection.

\subsection{Disaggregated Serving Systems}
% DistServe, LoongServe, TetriInfer, Splitwise, Mooncake, KVDirect, WindServe, semi-PD, MemServe
Traditional phase-wise unified systems typically co-locate both the prefill and decode phases within monolithic architectures, often leading to significant interference due to the divergent characteristics of these stages. Recent research has revealed fundamental differences between the prefill and decode phases, prompting the development of novel phase-disaggregated system designs. DistServe \cite{zhong2024distserve} proposes this approach through GPU-level disaggregation, dynamically allocating separate GPU devices for each phase while introducing parallelism adjustments to optimize resource utilization. Building on this, LoongServe \cite{wu2024loongserve} further refines the paradigm through sequence parallelism and flash decoding \cite{hong2024flashdecoding++}, significantly improving throughput for long-context scenarios. The trade-off for these compute-centric approaches lies in their operational complexity, as they require careful synchronization between disaggregated components and may incur overhead during phase transitions. TetriInfer \cite{hu2024tetriInfer} introduces a two-level scheduling mechanism enhanced with predictive resource profiling, effectively preventing decode-phase scheduling bottlenecks through anticipatory workload distribution.

As cluster sizes expand, disaggregated architectures are becoming increasingly popular in distributed scenarios, with recent work contributing to this area. Taking a more radical approach, Splitwise \cite{patel2024splitwise} advances system disaggregation by partitioning and distributing distinct processing phases across dedicated machines, rather than merely separating GPU devices. This architectural strategy facilitates phase-specific resource allocation, enabling each computation stage to leverage hardware optimally suited to its particular requirements. Mooncake \cite{qin2024mooncake} and MemServe \cite{hu2024memserve} explore distributed memory management approaches, with Mooncake building a KVCache-centric disaggregated architecture through a distributed KV Cache pool and scheduling based on different SLOs for clusters at various stages. Meanwhile, MemServe completes an elastic memory pool for managing memory across instances. Moreover, KVDirect \cite{chen2024kvdirect} proposes a tensor-centric communication mechanism and library for different machines, utilizing Remote Direct Memory Access (RDMA) technology to facilitate KV transmission between distributed machines.

Considering that stream-based disaggregated architectures are an emerging system design, WindServe \cite{feng2025windserve} proposes separating prefill and decode tasks onto different streams, utilizing fine-grained dynamic scheduling based on streams to enhance resource utilization and performance. In contrast, semi-PD \cite{hong2025semi} attempts to introduce Streaming Multiprocessor (SM)-level phase separation and further designs a dynamic partitioning algorithm based on SLO awareness to maximize the inherent parallelism of GPUs.

The existing disaggregated architectures face several drawbacks. One significant issue is the imbalance in resource utilization, particularly with the prefill phase, which often leads to wasted GPU memory. This inefficiency arises because the resources allocated for prefill may not be fully utilized, resulting in suboptimal performance. Additionally, the migration of requests between the prefill and decode phases can introduce stalls, further degrading overall system throughput and responsiveness. While distributed approaches emphasize the transmission of KV Cache data between machines or instances, they frequently overlook the impact of model reconfiguration. This oversight makes it challenging to handle sudden surges in requests, as the system may not adapt quickly enough to the changing workload demands. Moreover, stream-based methods impose specific hardware requirements that may limit their applicability across different environments, potentially restricting their scalability and flexibility.

\subsection{Load Balancing Scheduling}
% Llumnix, spotserve, uellm, 
Load balancing scheduling techniques are essential in disaggregated architectures, as both are aimed at reducing SLO violation rates. In modern distributed systems, effective load balancing is crucial for optimizing resource utilization and ensuring that all components operate efficiently. By strategically distributing workloads across various processing units or nodes, load balancing prevents any single unit from becoming a bottleneck, which is particularly important when different phases of processing, such as prefill and decode, are handled by separate units.

Intelligent routing of requests based on current load and resource availability helps maintain consistent performance and meet SLOs. Notable approaches in this domain include Llumnix \cite{Llumnix}, which focuses on dynamic load balancing by continuously monitoring the performance of different nodes and redistributing tasks as needed to maintain equilibrium. SpotServe \cite{SpotServe} takes a cost-effective approach by leveraging spot instances in cloud environments, optimizing resource allocation while managing the inherent volatility of these instances. UELLM \cite{uellm} implements an adaptive load balancing strategy that adjusts dynamically to changing workloads, enhancing the system's resilience and ability to handle unexpected spikes in demand.

However, existing load balancing strategies also have notable limitations when applied to prefill and decode disaggregated LLM serving. Most prior approaches are designed for homogeneous workloads and do not explicitly address the intrinsic compute and memory imbalance between prefill and decode stages. Moreover, they typically assume that workload distribution decisions can be made independently of state placement, overlooking the constraints imposed by large KV Cache locality. As a result, cache-aware routing may still lead to load skew, where high-hit-rate nodes become overloaded while other nodes remain underutilized. These gaps motivate the need for new scheduling mechanisms that are both state-aware and capable of dynamically rebalancing heterogeneous resources across stages.

% 画个表来总结一下每个work 和 banaserve的区别
\subsection{Summary}

Existing LLM serving frameworks, including vLLM, SGLang, DeepSpeed-FastGen, TensorRT-LLM, and NVIDIA Dynamo, focus on kernel-level optimizations, KV Cache management, and execution graph compilation within monolithic architectures. While these systems achieve substantial performance gains, their tightly coupled execution of the prefill and decode phases exacerbates resource contention and limits flexibility under dynamic workloads. Phase-disaggregated architectures, such as DistServe, LoongServe, TetriInfer, Splitwise, Mooncake, KVDirect, WindServe, semi-PD, and MemServe, mitigate prefill and decode interference through physical or logical separation of phases. However, they introduce new challenges, including load imbalance, resource fragmentation, inter-phase migration stalls, and scalability limitations under heterogeneous hardware conditions. Complementary research on load balancing scheduling, such as Llumnix, SpotServe, and UELLM, optimizes request distribution but typically targets generic distributed inference and does not specifically address the unique constraints of LLM phase separation and prefix caching.

BanaServe addresses two limitations that remain unresolved by prior work. First, it uses a unified KV Cache store to eliminate the load imbalance caused by prefix-caching-aware routing in the prefill phase, enabling routing decisions to be based solely on node load. Second, it introduces fine-grained module migration at the layer and attention computation levels, facilitating dynamic resource rebalancing between prefill and decode phases to adapt to fluctuating request patterns.

\section{System  Design}\label{sec 4}
% 动态的module的迁移 => 实现细粒度的pd之间的负载均衡
% 统一的KV Cache管理 => 调度器只需关注每个GPU的负载，无需关注prefix caching 的利用率
In this section, we present the architectural and algorithmic components of \textbf{BanaServe}, a LLM serving framework designed for PD disaggregated architecture in AI infrastructure. We first describe the Dynamic Migration mechanism for computational modules, which enables fine and coarse grained redistribution of workloads between GPUs in real time. We then introduce the Global KV Cache Store, a unified cache management layer that allows cross-instance KV Cache sharing and substantially simplifies scheduling logic. Building upon these designs, we formulate analytical models for PD disaggregation performance to guide system optimization. Finally, we detail two complementary runtime algorithms, Dynamic Migration and Load-aware Request Scheduling, which together maintain balanced utilization, minimize latency, and sustain high throughput under dynamic workload conditions.

\subsection{Dynamic Migration For Modules}

% 我们提出了两种不同粒度的迁移方式：层的迁移和attnetion的迁移（画两个图）
% 层的迁移：将模型的某些（尽量连续）的层迁移到其他的卡中。实现动态的模型并行。
% attention的迁移：将KV Cache按头进行切割和迁移，只迁移计算量并不会迁移权重，更加灵活。
To address the inherent inefficiencies of static PD disaggregation architectures, we propose a \textbf{dynamic resource allocation paradigm} based on module migration. This design enables the scheduler to adaptively redistribute computational modules, ranging from entire transformer blocks to portions of the KV Cache, across GPUs in response to real-time workload conditions. By dynamically relocating modules between prefill and decode instances, our framework achieves superior utilization of both computation and memory resources.

We formulate the migration decision as an optimization problem that aims to minimize the peak utilization across all devices:
\begin{equation}
\min_{\mathcal{M}} \; \max_{g \in \mathcal{G}} \; U_g(\mathcal{M}),
\end{equation}
where $\mathcal{G}$ is the set of available GPUs, $\mathcal{M}$ denotes the chosen module migration plan, and $U_g(\mathcal{M})$ is the utilization of GPU $g$ under $\mathcal{M}$. 

Migration feasibility is constrained by a per-orchestration budget on latency overhead $T_{\text{budget}}$:
\begin{equation}
T_{\text{mig}}(\mathcal{M}) \leq T_{\text{budget}},
\end{equation}
where $T_{\text{mig}}$ denotes the migration time, composed of \emph{weight transfer} and \emph{state transfer} latency.

\begin{figure}[t]
    \centering
    \includegraphics[width=0.65\linewidth]{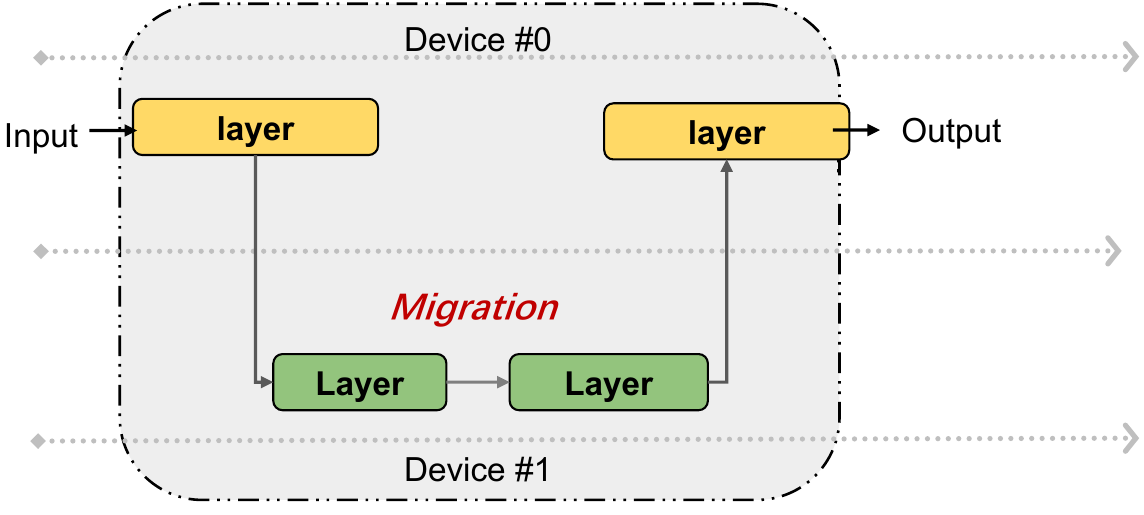}
    \caption{
    \textbf{Layer level migration in BanaServe:} A set of adjacent Transformer layers is dynamically reallocated across prefill and decode GPUs, with both layer weights $W_{\ell}$ and KV Cache $\mathcal{K}\mathcal{V}_{\ell}$ migrated.  
    Local execution resumes after payload transfer, enabling coarse grained compute and memory rebalancing while preserving output correctness.  
    Device \#0 and Device \#1 process different segments in parallel until migration completes, avoiding service disruption.
    }
    \label{fig:offload}
    \vspace{-1em}
\end{figure}

We design and implement two complementary migration granularities, providing a flexible trade-off between rebalancing precision and migration overhead:
%在这里增加更多的公式
% 1. layer的migration的迁移怎么进行？也可以画点图
% 2. 

\noindent\textbf{(1) Layer-level migration} (Fig.~\ref{fig:offload}):  
In scenarios where workload imbalance is pronounced, we migrate a contiguous set of Transformer layers from one GPU instance to another.  
This \emph{coarse grained} reallocation realizes \emph{dynamic model parallelism}, shifting both \textbf{layer weights} and the associated \textbf{KV Cache} to redistribute substantial compute and memory footprint between prefill and decode stages.

\textbf{Data volume to migrate:}  
Let $S_{\ell}^{\text{w}}$ be the size of the migrated layer weights and  
$S_{\ell}^{\text{kv}}$ be the size of the corresponding KV Cache state.  
The total payload is:
\begin{equation}
S_{\ell}^{\text{total}} = S_{\ell}^{\text{w}} + S_{\ell}^{\text{kv}}.
\end{equation}

\textbf{Latency model:}  
Given effective interconnect bandwidth $B_{\text{net}}$ and synchronization overhead $T_{\text{sync}}$,  
the migration latency is approximated by:
\begin{equation}
T_{\text{layer}} \approx \frac{S_{\ell}^{\text{total}}}{B_{\text{net}}} + T_{\text{sync}}.
\end{equation}
Since $S_{\ell}^{\text{w}} \gg S_{\ell}^{\text{kv}}$ in most cases,  
the latency is dominated by weight transfer, but still remains practical within high bandwidth 
data center fabrics (NVLink, Infiniband).

\textbf{Execution correctness:}  
For a migrated layer $f_{\ell}(\cdot)$ with weights $W_{\ell}$ and KV Cache $\mathcal{K}\mathcal{V}_{\ell}$,
the same computation can be carried out on the new GPU without semantic change:
\begin{equation}
\mathbf{y}_{\ell} = f_{\ell}(\mathbf{x}_{\ell} ; W_{\ell}, \mathcal{K}\mathcal{V}_{\ell}),
\end{equation}
where $\mathbf{x}_{\ell}$ is the input activation from the preceding layer.  
By transferring $(W_{\ell}, \mathcal{K}\mathcal{V}_{\ell})$ together, we ensure that 
auto-regressive decoding continues seamlessly after migration.

While this method incurs temporary weight relocation cost, it can rapidly mitigate severe imbalance by shifting large computational segments,  
and is particularly effective in bursty or highly skewed workloads.

% \noindent \textbf{(1) Layer-level migration} (Fig.~\ref{fig:offload}):  
% In scenarios where workload imbalance is pronounced, we migrate a set of transformer layers—preferably contiguous in depth—from one instance to another GPU. This \emph{coarse-grained} reallocation effectively realizes \emph{dynamic model parallelism}, redistributing substantial compute and memory footprint between the prefill and decode stages.  
% Let $S_{\ell}$ denote the size of parameters in migrated layers, and $B_{\text{net}}$ the effective interconnect bandwidth; the migration latency is approximated by:
% \begin{equation}
% T_{\text{layer}} \approx \frac{S_{\ell}}{B_{\text{net}}} + T_{\text{sync}}
% \end{equation}
% where $T_{\text{sync}}$ accounts for synchronization with ongoing computation.
% While this method incurs temporary weight relocation cost, it can rapidly mitigate severe imbalance by shifting large computational segments.
% \begin{figure}
%     \centering
%     \includegraphics[width=0.65\linewidth]{Properties/offload.pdf}
%     \caption{Layer-level migration in BanaServe: adjacent Transformer layers (blocks) are dynamically reallocated across prefill and decode GPUs to address coarse-grained compute and memory imbalance. This mechanism enables adaptive model-parallel reconfiguration without service disruption.}
%     \label{fig:offload}
%     \vspace{-1em}
% \end{figure}

\begin{figure}[t]
    \centering
    \includegraphics[width=0.85\linewidth]{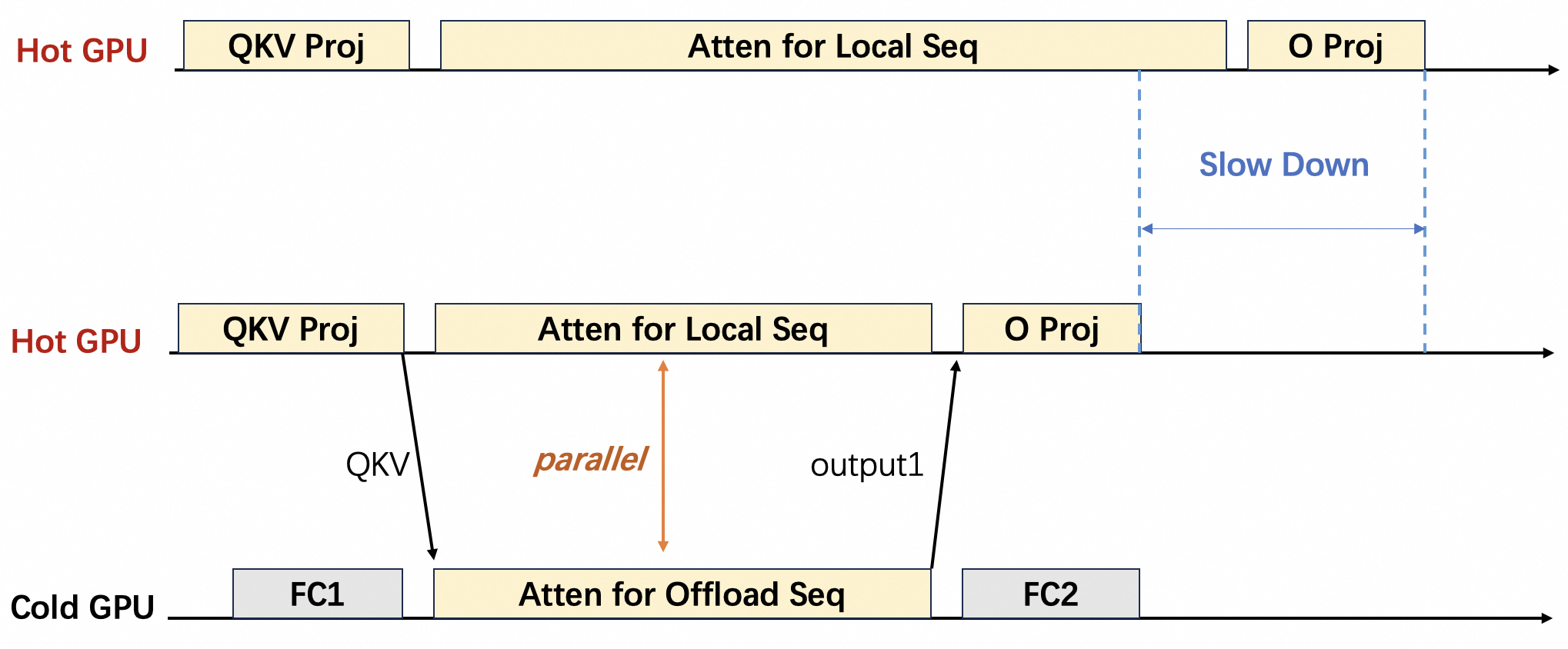}
    \caption{
    \textbf{Attention level migration in BanaServe.}
    KV Cache is split along the attention head dimension into $K^{(1)},V^{(1)}$ (hot GPU) and $K^{(2)},V^{(2)}$ (cold GPU).
    Local attention (\textit{Atten for Local Seq}) and offloaded attention (\textit{Atten for Offload Seq}, preceded by FC1 and followed by FC2) are computed in parallel.
    Only the partial softmax denominator $\ell^{(1)}$ and output $O^{(1)}$ are exchanged across devices, enabling minimal data transfer and mitigating hotspots with negligible service disruption.
    }
    \label{fig:attention_offload}
    \vspace{-1em}
\end{figure}

\noindent\textbf{(2) Attention-level migration} (Fig.~\ref{fig:attention_offload}):  
When fine-grained load adjustment is required, we partition the KV Cache along the attention head dimension and selectively migrate only the KV states of certain heads to auxiliary \textit{cold} GPUs. Since no model weights are transferred and only intermediate key and value activations are moved, this method imposes minimal migration overhead.

% --- 图解说明 ---
The pipeline begins with the \textit{QKV Projection} layer, which generates query ($Q$), key ($K$), and value ($V$) matrices from the input hidden states.
We split the attention heads into two disjoint subsets:
\begin{itemize}
    \item $K^{(1)}, V^{(1)}$ (\textbf{KV$_1$}): retained on the hot GPU for local processing.
    \item $K^{(2)}, V^{(2)}$ (\textbf{KV$_2$}): offloaded to a cold GPU for remote processing.
\end{itemize}
The local device executes \textit{Atten for Local Seq} using $K^{(1)}, V^{(1)}$, while the cold device executes \textit{Atten for Offload Seq} using $K^{(2)}, V^{(2)}$ in parallel.
Blocks \textbf{FC1} and \textbf{FC2} are lightweight fully connected layers bracketing the offloaded attention on the cold GPU, and \textit{O Proj} is the output projection of the multi head attention (MHA).

% --- 并行可行性公式 ---
\textbf{Parallel computation feasibility:}
Let $H$ be the total number of attention heads, $d$ the head dimension,  
$Q \in \mathbb{R}^{B_q \times Hd}$ the query matrix,  
and $K^{(j)}, V^{(j)} \in \mathbb{R}^{B_k \times d_{h_j}}$ 
the key/value subsets assigned to device $j \in \{1,2\}$.
The attention score and value aggregation for each subset are:
\begin{align}
S^{(j)} &= Q \cdot (K^{(j)})^\top, \\
A^{(j)} &= \exp\left(S^{(j)}\right), \\
\ell^{(j)} &= 
    \begin{cases}
        \sum_i A^{(1)}_i, & j = 1, \\
        \ell^{(1)} + \sum_i A^{(2)}_i, & j = 2,
    \end{cases} \\
O^{(j)} &= \frac{A^{(j)}}{\ell^{(j)}} \cdot V^{(j)}.
\end{align}
The final attention output is:
\begin{equation}
O = O^{(1)} + O^{(2)}.
\end{equation}

Since $K^{(1)}, V^{(1)}$ and $K^{(2)}, V^{(2)}$ are from disjoint head partitions,  
$S^{(1)}$ and $S^{(2)}$ can be computed entirely in parallel on separate devices without intermediate data dependencies.  
Only $\ell^{(1)}$ needs to be communicated to the cold GPU before its normalization step,  
and $O^{(1)}$ (or $\ell^{(2)}$) is sent back before the final summation,  
ensuring correctness of the global softmax while keeping transfer size minimal.

% --- 迁移延迟公式 ---
Let $S_{\text{kv}}$ denote the size of migrated KV data and $B_{\text{net}}$ is the  network bandwidth, thus the migration latency is:
\begin{equation}
T_{\text{attn}} \approx \frac{S_{\text{kv}}}{B_{\text{net}}}.
\end{equation}
Typically $S_{\text{kv}} \ll S_\ell$ (layerwise state size), hence $T_{\text{attn}} \ll T_{\text{layer}}$.  
This makes attention level migration lightweight, latency friendly, and effective in real time load rebalancing for heterogeneous decoding workloads.

% --- 图注优化 ---

\noindent
By combining coarse grained layer migration with fine grained attention migration,  
BanaServe can adaptively select at runtime the strategy that balances migration cost ($T_{\text{mig}}$) against the expected gain in utilization (minimizing $\max_g U_g$),  
under latency budget constraint $T_{\text{mig}} \leq T_{\text{budget}}$.

\subsection{Global KV Cache Store}
% prefix cache aware router造成prefill节点间的负载不均衡的本质原因是prefill节点间的存储的不均衡，prefill节点间无法共享最大的perfix cache，因此我们在prefill和decode节点间使用了全局的KVCache，所有的Prefill节点共享KV Cache，这也就意味着prefix cache aware router 只需要关注prefill节点的负载即可，prefix cache的load和fetch只需要交给KV Cache Store即可。
% KV Cache Store 的一大难点是load和fetch产生的延迟很高，我们根据layer wised 的特性，设计了三级流水，将pre fetch ，计算，load进行overlap，极大的降低了延迟，进而可以实现无感的prefill计算。
The primary bottleneck of the prefix cache aware router lies in the imbalance of prefix cache storage across prefill nodes. Since the largest prefix cache segment cannot be shared among different prefill nodes, the router must consider both computation and cache placement when dispatching requests. To address this problem, we introduce a \textbf{Global KV Cache Store} that spans both prefill and decode nodes. With this design, all prefill nodes access a shared KV Cache, enabling the router to focus solely on balancing computational workload. All prefix cache \texttt{load} and \texttt{fetch} operations are delegated to the KV Cache Store.
\begin{figure}
    \centering
    \includegraphics[width=0.65\linewidth]{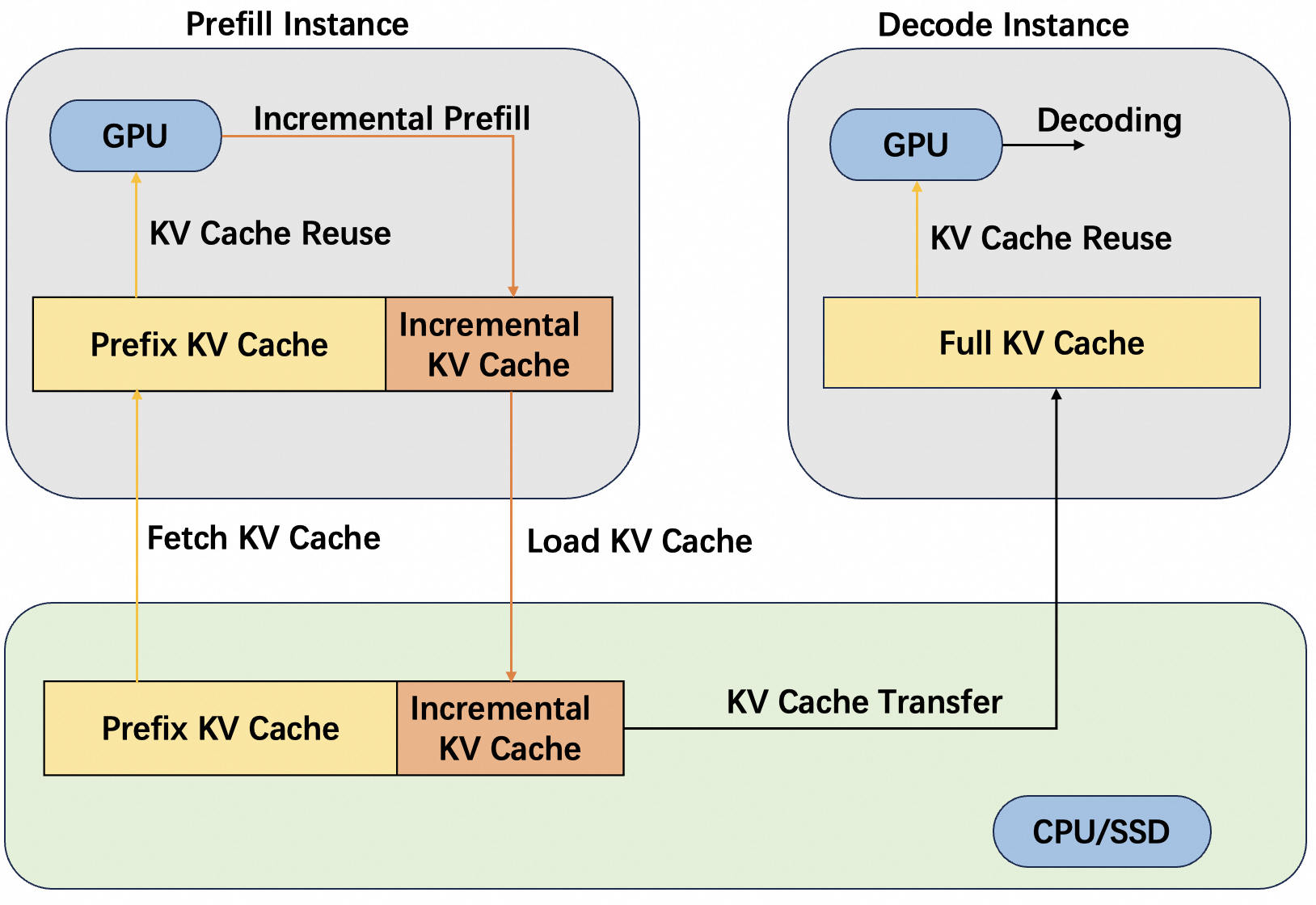}
    \caption{KV Cache management and reuse in BanaServe. Prefill instances perform incremental prefill on incoming requests, reusing cached prefixes when available. Prefix KV Cache and newly generated incremental KV Cache are stored in a shared CPU/SSD-backed KV store, enabling both cache sharing across prefill GPUs and complete KV Cache assembly. During decoding, decode instances retrieve the full KV Cache from the shared store and reuse it for token generation, avoiding redundant computation and supporting efficient prefill and decode execution.}
    \label{fig:kv store}
    \vspace{-1em}
\end{figure}

A major challenge in implementing a Global KV Cache Store is the latency incurred by repeated \texttt{load} and \texttt{fetch} operations. To mitigate this, we leverage the \textbf{layer-wise} execution property of transformer models to design a \textbf{three-stage pipeline} where prefetching, computation, and loading are overlapped. This overlap hides most of the memory access latency, resulting in near-transparent prefill computation without observable throughput degradation.

We let \(T_{\mathrm{F}}\) denote the total forward computation time for the prefill phase, 
\(r\) is the average prefix cache hit rate, 
\(L\) is the input sequence length in tokens, 
and \(N\) denotes the total number of Transformer layers.  
The per-layer forward computation time is then
\begin{equation}
T_{\mathrm{F,layer}} = \frac{T_{\mathrm{F}} \cdot r}{N}.
\end{equation}
Similarly, let \(S_{\mathrm{kv}}\) be the per-layer KV Cache size for a single token (in bytes), 
and \(B\) is the effective PCIe bandwidth.  
The per-layer KV Cache transfer time is
\begin{equation}
T_{\mathrm{KV}} = \frac{S_{\mathrm{kv}} \cdot L \cdot r}{B}.
\end{equation}

For example, in a \texttt{llama-3.1-8B} model with hidden size \(d_{\mathrm{model}} = 4096\), 
total attention heads \(h_{\mathrm{total}} = 32\), 
KV heads under GQA \(h_{\mathrm{kv}} = 8\), 
and BF16 precision (\(2\) bytes per element), 
the dimension per head is
\begin{equation}
d_{\mathrm{head}} = \frac{d_{\mathrm{model}}}{h_{\mathrm{total}}} = 128.
\end{equation}

Because each KV head stores both \emph{key} and \emph{value} activations,  
the per-layer KV Cache size is
\begin{equation}
S_{\mathrm{kv}} = h_{\mathrm{kv}} \cdot d_{\mathrm{head}} \cdot 2 \cdot 2\ \mathrm{bytes} 
= 8 \times 128 \times 2 \times 2\ \mathrm{bytes} = 4096\ \mathrm{bytes} \ (4\,\mathrm{KB}).
\end{equation}
Multiplying by \(N=32\) layers gives the total KV Cache per token:
\begin{equation}
S_{\mathrm{kv,total}} = 32 \times 4\,\mathrm{KB} = 128\,\mathrm{KB}.
\end{equation}

In our evaluation with \(L = 1000\) tokens, \(r = 0.5\), \(B = 200\ \mathrm{Gbps}\), and \(T_{\mathrm{F}} = 270\ \mathrm{ms}\), we have:
\begin{align}
T_{\mathrm{F,layer}} = \frac{270\ \mathrm{ms} \times 0.5}{32} \approx 4.22\ \mathrm{ms}, 
\quad
T_{\mathrm{KV}} = \frac{4\ \mathrm{KB} \times 1000 \times 0.5}{200\ \mathrm{Gbps}} \approx 0.082\ \mathrm{ms}.
\end{align}
Since \(T_{\mathrm{KV}} \ll T_{\mathrm{F,layer}}\), the KV Cache transfer latency is negligible compared with computation time, enabling full overlap of communication and computation in the proposed pipeline. This overlap ensures that KV Cache fetch and store operations remain transparent to the serving workflow, sustaining high throughput without observable stalls in the prefill phase.

% Under a representative hardware setup (NVIDIA RTX~4090 with 24\,GB memory, PCIe~4.0$\times$16 at 200\,Gbps), a layer-wise KV Cache size of 4\,KB, full KV Cache size of 128\,KB, input sequence length of 1000 tokens, output length of 100 tokens, and an average prefix cache hit rate of 50\%, the measured per-layer forward time is:
% \begin{equation}
% T_{\text{FWD-layer}} = \frac{270\text{ ms} \times 50\%}{1000} \approx 0.14\text{ ms}
% \end{equation}
% while the per-layer KV Cache transfer time is:
% \begin{equation}
% T_{\text{KV-transfer}} = \frac{4\text{ KB} \times 1000 \times 50\%}{200\text{ Gbps}} \approx 0.082\text{ ms} \ll T_{\text{FWD-layer}}
% \end{equation}
% Since the transfer time is significantly smaller than the compute time, the pipeline can fully overlap KV Cache communication with computation.

\begin{figure}
    \centering
    \includegraphics[width=0.85\linewidth]{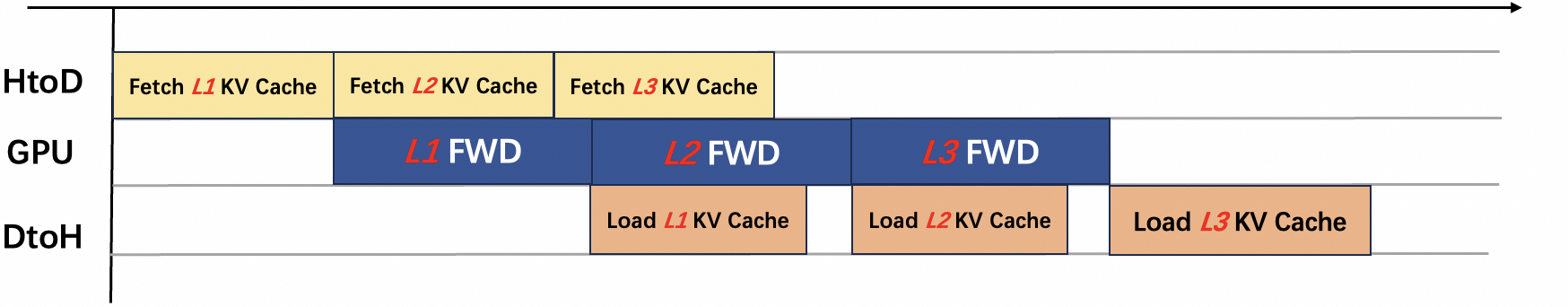}
    \caption{Validation of the three-stage layer-wise KV Cache pipeline in BanaServe. The top section lists hardware parameters and measured latencies for a representative workload. Under a 50\% average prefix cache hit rate, the per-layer forward time ($4.22$\,ms) exceeds the per-layer KV Cache transfer time ($0.082$\,ms), ensuring effective overlap. The timeline shows the execution for three layers ($L1$–$L3$): while the GPU executes forward computation for $L_i$, the host-to-device (HtoD) channel fetches KV Cache for $L_{i+1}$ and the device-to-host (DtoH) channel stores cache from $L_{i-1}$. This concurrency hides communication latency and enables transparent prefill execution.}
    \label{fig:kv transfer}
    \vspace{-1em}
\end{figure}

This analytic and empirical validation confirms that the three-stage pipeline design is effective in overlapping cache transfer with computation. Consequently, the Global KV Cache Store can serve prefix cache requests without introducing noticeable stalls in prefill execution, which is critical for sustaining high throughput in PD disaggregated architectures.

\subsection{Modeling PD Disaggregation Performance}
We formulate the performance optimization in PD disaggregation systems as a multi-objective problem, aiming to \emph{maximize resource utilization}, \emph{minimize end-to-end latency}, and thereby \emph{maximize throughput} across both prefill and decode stages.

Let $\mathcal{G}_p$ and $\mathcal{G}_d$ denote the sets of prefill and decode GPUs respectively, and $U_g$ the utilization of GPU $g$.  
Let $T_{\text{TTFT}}$ denote the time to first token and $T_{\text{TPOT}}$ is the time per output token.  
Let $\Theta$ denote the effective system throughput (tokens/sec).

\subsubsection{Optimization Objective}
We define the joint optimization objective as:
\begin{equation}
\max_{\mathcal{M}} \;\; \alpha \cdot U_{\text{avg}}(\mathcal{M}) \;-\; \beta \cdot T_{\text{avg-latency}}(\mathcal{M}) \;+\; \gamma \cdot \Theta(\mathcal{M}),
\end{equation}
where:
\begin{equation}
U_{\text{avg}}(\mathcal{M}) = \frac{1}{|\mathcal{G}_p \cup \mathcal{G}_d|} \sum_{g} U_g(\mathcal{M}),
\quad
T_{\text{avg-latency}} = \frac{1}{N} \sum_{i=1}^N T_{\text{TTFT}}^{(i)} + T_{\text{TPOT}}^{(i)},
\end{equation}
\(\alpha, \beta, \gamma\) are weighting coefficients for utilization, latency, and throughput; $\mathcal{M}$ denotes the current module migration plan.

\subsubsection{Latency Model}
The Time to First Token (TTFT) is modeled as:
\begin{equation}
T_{\mathrm{TTFT}} = T_{\mathrm{p}} + T_{\mathrm{x}} + T_{\mathrm{q}},
\end{equation}
where \(T_{\mathrm{p}}\) is the prefill computation time,  
\(T_{\mathrm{x}}\) is the KV Cache transfer time,  
and \(T_{\mathrm{q}}\) is the queuing delay before decode.
The KV transfer time is decomposed as
\begin{equation}
T_{\mathrm{x}} = T_{\mathrm{load}} + T_{\mathrm{fetch}},
\end{equation}
with \(T_{\mathrm{load}}\) denoting the time to load KV state from storage and \(T_{\mathrm{fetch}}\) means the time to fetch KV state from remote GPUs.

For the decode phase, the time per output token (TPOT) is:
\begin{equation}
T_{\mathrm{TPOT}} = T_{\mathrm{d}} + T_{\mathrm{c}} + T_{\mathrm{m}},
\end{equation}
where \(T_{\mathrm{d}}\) is the base decoding compute time,  
\(T_{\mathrm{c}}\) is the cache access latency,  
and \(T_{\mathrm{m}}\) is the memory bandwidth stall time.

\subsubsection{Resource Utilization Model}
Consider a prefill instance with \(n_p\) Transformer layers.  
Let \(M_{0}\) be the base memory overhead of the process,  
\(M_{\ell}\) is the memory per layer,  
and \(K_{\mathrm{init}}\) is the initial KV Cache allocation.  
The total memory footprint is
\begin{equation}
\mathrm{Mem}_{p} = M_{0} + n_p \cdot M_{\ell} + K_{\mathrm{init}}.
\end{equation}

Let \(C_{\ell}\) be the compute cost per layer per token,  
\(B_{\mathrm{sz}}\) is the batch size,  
and \(L_{\mathrm{in}}\) denotes the input length in tokens.  
The total compute demand is
\begin{equation}
\mathrm{Comp}_{p} = n_p \cdot C_{\ell} \cdot B_{\mathrm{sz}} \cdot L_{\mathrm{in}}.
\end{equation}

Similarly, for a decode instance with \(n_d\) layers:  
let \(K_{\mathrm{acc}}\) be the KV Cache size accumulated during decoding,  
and \(L_{\mathrm{gen}}\) is the generation length in tokens.  
We have:
\begin{align}
\mathrm{Mem}_{d} &= M_{0} + n_d \cdot M_{\ell} + K_{\mathrm{acc}}, \\
\mathrm{Comp}_{d} &= n_d \cdot C_{\ell} \cdot B_{\mathrm{sz}} \cdot L_{\mathrm{gen}}.
\end{align}

Given peak GPU compute capacity \(C_{\mathrm{gpu}}\),  
the average utilization of each stage is:
\begin{equation}
U_{p} = \frac{\mathrm{Comp}_{p}}{C_{\mathrm{gpu}}}, \quad
U_{d} = \frac{\mathrm{Comp}_{d}}{C_{\mathrm{gpu}}}.
\end{equation}

\subsubsection{Migration Cost}
Migrating \(k\) modules between instances incurs total overhead:
\begin{equation}
\mathrm{Cost}_{\mathrm{mig}} = k \left( T_{\mathrm{x\_lat}} + T_{\mathrm{sync}} + T_{\mathrm{mem\_realloc}} \right),
\end{equation}
where \(T_{\mathrm{x\_lat}}\) is the KV+weight transfer latency,  
\(T_{\mathrm{sync}}\) represents the synchronization time with ongoing computation,  
and \(T_{\mathrm{mem\_realloc}}\) denotes the time to reallocate buffers.

Module migration \(\mathcal{M}\) must satisfy latency constraints:
\begin{equation}
T_{\mathrm{TTFT}}(\mathcal{M}) \leq T_{\mathrm{TTFT}}^{\mathrm{budget}}, \quad
T_{\mathrm{TPOT}}(\mathcal{M}) \leq T_{\mathrm{TPOT}}^{\mathrm{budget}}.
\end{equation}

\subsubsection{Throughput}
Given \(N\) concurrent requests and total output length \(L_{\mathrm{out}}\) (tokens) per request, the system throughput \(\Theta\) is

% Given $N$ concurrent requests and total output length $L_{\text{out}}$ per request:
\begin{equation}
    \Theta = \frac{N \cdot L_{\text{out}}}{T_{\text{TTFT}} + L_{\text{out}} \cdot T_{\text{TPOT}}},
\end{equation}

and the scheduler periodically selects $\mathcal{M}^*$ to maximize the joint objective:
\begin{equation}
\mathcal{M}^* = \arg\max_{\mathcal{M}} \left( \alpha U_{\text{avg}}(\mathcal{M}) - \beta T_{\text{avg\_latency}}(\mathcal{M}) + \gamma \Theta(\mathcal{M}) \right),
\end{equation}
where \(\alpha, \beta, \gamma\) are weighting coefficients corresponding to the utilization, latency, and throughput terms in the objective function. 
subject to migration cost and latency constraints.

% To formalize the optimization problem, we develop analytical models for key performance metrics in PD separation systems. The Time to First Token (TTFT) can be modeled as:
% $$\text{TTFT} = T_{\text{prefill}} + T_{\text{transfer}} + T_{\text{queue\_decode}}$$

% The Time Per Output Token (TPOT) in the decode phase follows:
% $$\text{TPOT} = T_{\text{decode\_base}} + T_{\text{cache\_access}} + T_{\text{memory\_bandwidth}}$$

% Resource utilization for prefill and decode instances can be modeled as functions of allocated modules. For a prefill instance with $n_p$ layers:
% \begin{align}
% \text{Memory}_{\text{prefill}} &= \text{Base}_{\text{memory}} + n_p \times \text{Layer}_{\text{memory}} + \text{KV}_{\text{cache\_init}} \\
% \text{Compute}_{\text{prefill}} &= n_p \times \text{Layer}_{\text{compute}} \times \text{Batch}_{\text{size}} \times \text{Sequence}_{\text{length}}
% \end{align}

% Similarly, for decode instances with $n_d$ layers:
% \begin{align}
% \text{Memory}_{\text{decode}} &= \text{Base}_{\text{memory}} + n_d \times \text{Layer}_{\text{memory}} + \text{KV}_{\text{cache\_accumulated}} \\
% \text{Compute}_{\text{decode}} &= n_d \times \text{Layer}_{\text{compute}} \times \text{Batch}_{\text{size}} \times \text{Generation}_{\text{length}}
% \end{align}

% The migration of $k$ modules between instances incurs overhead costs modeled as:
% $$\text{Migration}_{\text{cost}} = k \times (\text{Transfer}_{\text{latency}} + \text{Synchronization}_{\text{overhead}} + \text{Memory}_{\text{reallocation}})$$

\subsection{Adaptive Migration and Load-aware Scheduling in BanaServe}
Based on our performance models and experimental observations, BanaServe employs two complementary runtime algorithms: 
\begin{enumerate}
    \item a \textbf{Dynamic Migration Algorithm }that adaptively rebalances computation and memory between prefill and decode instances through module migration.
    \item a \textbf{Load-aware Request Scheduling Algorithm} that dispatches requests solely based on instance load, enabled by a Global KV Cache Store.
\end{enumerate}

\subsubsection{Dynamic Migration Algorithm}
\begin{algorithm}[htbp]
\caption{Adaptive Module Migration Algorithm in BanaServe}
\label{alg:migration}
\begin{algorithmic}[1]
\State \textbf{Input:} Set of devices $D$, current module allocation $cfg$, balance threshold $\delta$
\State \textbf{Output:} Updated allocation $cfg'$

\State // Step 1: Measure current load
\For{each $d$ in $D$}
    \State $load[d] \gets \text{MeasureUtilization}(d)$ \Comment{Compute+Memory}
\EndFor
\State $overload \gets \{d \in D \mid load[d] - \min(load) > \delta \}$ 
\State $underload \gets \{d \in D \mid \max(load) - load[d] > \delta \}$

\State // Step 2: Migration decision loop
\While{$\exists\, d_o \in overload$ and $\exists\, d_u \in underload$}
    \State $\Delta \gets load[d_o] - load[d_u]$
    \If{$\Delta \geq \delta\ \land\ \text{SupportsLayerMigration}(d_o)$}
        \State $\text{MigrateLayer}(d_o, d_u)$
    \ElsIf{$\Delta \geq \delta\ \land\ \text{SupportsAttentionMigration}(d_o)$}
        \State $\text{MigrateKVHeads}(d_o, d_u)$
    \EndIf
    \State Update $load[d_o], load[d_u]$
    \State Update $cfg'$ accordingly
\EndWhile

\State \Return $cfg'$
\end{algorithmic}
\end{algorithm}

% \begin{algorithm}[htbp]
% \caption{Migration Algorithm}
% \label{alg:migration}
% \begin{algorithmic}[1]
% \State \textbf{Input:} Models $M$, Device $D$, Allocation config $cfg$, Balance threshold $\delta$
% \State \textbf{Output:} Updated config $cfg$
% \BlankLine
% \State Initialize $min\_load \gets 0$, $max\_load \gets 0$;
% \For{each $m$ \textbf{in} $M$}
%     \State Update $min\_load$ and $max\_load$ using $m.work\_load$;
%     \If{$max\_load - min\_load \leq \delta$} \Comment{Imbalance Condition}
%         \State $overload\_device \gets \text{find device with max load}$;
%         \State $underload\_device \gets \text{find device with min load}$;
%         \State Migrate a model from $overload\_device$ to $underload\_device$;
%         \State Update $cfg$ accordingly;
%     \EndIf
% \EndFor
% \State \Return $cfg$
% \end{algorithmic}
% \end{algorithm}

BanaServe uses a Dynamic Migration Algorithm to resolve compute and memory utilization imbalance between prefill and decode instances in PD disaggregation. The algorithm runs in a periodic control cycle that monitors real-time load and performs module migration to balance the workload while limiting migration overhead.

Let $D = \{d_1, d_2, \dots, d_{|D|}\}$ be the set of devices (GPUs) currently participating in serving, $cfg$ the current allocation of model modules, and $\delta > 0$ the load imbalance threshold. For each device $d$, we define its normalized utilization as
\begin{equation}
U_d = \frac{C_d}{C_d^{\max}} + \frac{M_d}{M_d^{\max}},
\end{equation}
where $C_d$ is the current compute usage, $M_d$ is the current memory usage, and $C_d^{\max}, M_d^{\max}$ are the respective hardware capacities. $U_d$ ranges from $0$ to $2.0$, with larger values indicating heavier combined compute and memory load.

The algorithm operates in four sequential stages that directly correspond to the steps in Algorithm~\ref{alg:migration}:

\begin{enumerate}
    \item[1] \textbf{Load measurement} (lines 3–5): For each device $d \in D$, compute the combined compute and memory utilization $load[d]$ by invoking \texttt{MeasureUtilization}. This provides each device’s current normalized load, which will be used to determine imbalance.
    
    \item[2] \textbf{Load classification} (lines 6–7): Identify overloaded and underloaded devices by constructing:
    \begin{equation}
        overload = \{d \in D \mid load[d] - \min(load) > \delta\}, \quad
        underload = \{d \in D \mid \max(load) - load[d] > \delta\}.
    \end{equation}

    Devices in \(overload\) exceed the lightest‑loaded device’s utilization by more than the threshold \(\delta\), while devices in \(underload\) fall below the heaviest‑loaded device’s utilization by more than \(\delta\).
    
    \item[3] \textbf{Migration decision and execution} (lines 9–17): While there exists $d_o \in overload$ and $d_u \in underload$, calculate the instantaneous load gap:
    \begin{equation}
        \Delta = load[d_o] - load[d_u].
    \end{equation}
    
    If \(\Delta \ge \delta\) and $d_o$ supports layer‑level migration, invoke \texttt{MigrateLayer}($d_o, d_u$) to transfer contiguous transformer layers.  
    If \(\Delta \ge \delta\) and $d_o$ supports attention‑level migration, invoke \texttt{MigrateKVHeads}($d_o, d_u$) to transfer selected KV Cache segments along attention heads.  
    Before execution, evaluate the migration using:
    \begin{equation}
    Benefit(m) = \Delta_{\text{before}} - \Delta_{\text{after}}, \quad
    Cost(m) = T_{\text{transfer}}(m) + T_{\text{sync}}(m),
    \end{equation}
    where $T_{\text{transfer}}(m)$ is the data transfer latency and $T_{\text{sync}}(m)$ is the synchronization cost with ongoing tasks. The operation proceeds only if \(Benefit(m) / Cost(m) \ge \rho\), where \(\rho\) is a configurable efficiency ratio.
    
    \item[4] \textbf{Update allocation} (lines 18–20): After each migration action, update $load[d_o]$ and $load[d_u]$, revise \(cfg'\) to reflect the new module placement, and repeat until no device simultaneously meets overload and underload criteria. Finally, return the updated \(cfg'\).
\end{enumerate}

The complexity per control cycle is
\begin{equation}
O(|D| + N_m),
\end{equation}
where $|D|$ is the number of devices and $N_m$ is the number of modules on the overloaded device that can be migrated. The $O(|D|)$ term comes from load measurement and classification, and $O(N_m)$ from module selection and cost evaluation. In typical deployments, $|D|$ is in the tens and $N_m$ is bounded by model depth, enabling real-time operation.

To prevent oscillations, hysteresis control is applied using distinct upward and downward thresholds $(\delta_{\uparrow}, \delta_{\downarrow})$. Integration with the Global KV Cache Store ensures that all KV states are preserved during migration, avoiding recomputation and maintaining steady throughput. This dynamic migration mechanism allows BanaServe to adaptively balance workloads using both coarse and fine-grained reallocation strategies under highly dynamic traffic patterns.

\subsubsection{Load-aware Request Scheduling Algorithm}
\begin{algorithm}[htbp]
\caption{Load-aware Request Scheduling Algorithm}
\label{alg:load-aware-scheduling}
\begin{algorithmic}[1]
\State \textbf{Input:} Request queue $Q$, Prefill instance set $P$, Load threshold $\delta_L$
\State \textbf{Output:} Dispatch plan $D^*$
\State // Step 1: Measure current load on each prefill instance
\For{each $p_i \in P$}
    \State $load[p_i] \gets \texttt{MeasureUtilization}(p_i)$ \Comment{Compute + Memory}
    \State $q\_len[p_i] \gets \texttt{GetQueueLength}(p_i)$
\EndFor
\State // Step 2: Sort instances by load and queue length
\State $candidates \gets \texttt{Sort}(P, key=(load, q\_len), order=\texttt{ascending})$
\State // Step 3: Dispatch requests to least-loaded instance
\For{each $req \in Q$}
    \State $p_{target} \gets \texttt{Select}(candidates, policy=\texttt{least-loaded})$
    \If{$load[p_{target}] < \delta_L$}
        \State \texttt{AssignRequest}($req, p_{target}$)
        \State $load[p_{target}] \gets load[p_{target}] + \texttt{EstimateLoad}(req)$
    \Else
        \State $p_{target} \gets \texttt{Select}(candidates, policy=\texttt{lowest-queue})$
        \State \texttt{AssignRequest}($req, p_{target}$)
    \EndIf
\EndFor
\State // Step 4: Return final dispatch plan
\State $D^* \gets$ \texttt{GetDispatchMap}(P)
\State \Return $D^*$
\end{algorithmic}
\end{algorithm}

% \begin{algorithm}[htbp]
% \caption{Queue Algorithm}
% \label{alg:scale-up}
% \begin{algorithmic}[1]
% \State \textbf{Input:} LLM $model$, Cluster $G$, Current state $P$, Configuration Coefficient $\gamma$, Replica Size $r$, Layer Num $n$
% \State \textbf{Output:} Optimal strategy $P^*$
% \BlankLine
% \State $sp\_best \gets \frac{1}{\gamma + \frac{(1-\gamma)}{n} \times \left\|\mathbf{1} \oslash P\right\|_1}$
% \For{each $g_{dst}$ \textbf{in} \texttt{GetEligibleNodes}($G$)}
%     \State $max\_replicas \gets g_{dst}.available / r$
%     \State $candidates \gets$ \texttt{SortCandidatesByContinuity}($P^*, g_{dst}, max\_replicas$)
%     \For{each $layer\_id$ \textbf{in} $candidates$}
%         \State $P' \gets P^*.\texttt{copy}()$
%         \State $P'.\texttt{addReplica}(layer\_id, g_{dst})$
%         \State $sp \gets \frac{1}{\gamma + \frac{(1-\gamma)}{n} \times \left\|\mathbf{1} \oslash P'\right\|_1}$
%         \If{$sp > sp\_best$}
%             \State \texttt{replicate}($model, layer\_id, g_{dst}$)
%             \State $sp\_best \gets sp$
%             \State $P^* \gets P'$
%         \EndIf
%     \EndFor
% \EndFor
% \State \Return $P^*$
% \end{algorithmic}
% \end{algorithm}

BanaServe employs a Load-aware Request Scheduling Algorithm to efficiently distribute incoming requests among prefill instances. Unlike cache aware routing policies, this scheduler does not need to consider prefix cache hit rate due to the Global KV Cache Store, which enables KV Cache sharing across all prefill GPUs. The scheduling decision is therefore based solely on real-time load metrics, ensuring balanced resource utilization and minimizing queuing delays.

Let $Q = \{q_1, q_2, \dots, q_{|Q|}\}$ denote the incoming request queue, $P=\{p_1, p_2, \dots, p_{|P|}\}$ denote the set of active prefill instances, and $\delta_L>0$ is the load threshold used to trigger alternate routing logic. For each prefill instance $p_i$, the normalized load is defined as
\begin{equation}
U_{p_i} = \frac{C_{p_i}}{C^{\max}_{p_i}} + \frac{M_{p_i}}{M^{\max}_{p_i}},
\end{equation}
where $C_{p_i}$ and $M_{p_i}$ represent the current compute and memory usage, and $C^{\max}_{p_i}$, $M^{\max}_{p_i}$ are the respective capacities.

The algorithm executes the following stages, with each step directly mapped to the corresponding lines in Algorithm~\ref{alg:load-aware-scheduling}:

\begin{enumerate}
    \item[1] \textbf{Load measurement} (lines 3–6): For each $p_i\in P$, measure $U_{p_i}$ using \texttt{MeasureUtilization} and record its queue length $q\_len[p_i]$ via \texttt{GetQueueLength}.
    \item[2] \textbf{Sorting} (line 8–9): Sort all instances in ascending order by load and then by queue length. This produces the candidate list for dispatch decisions.
    \item[3] \textbf{Dispatch loop} (lines 11–19): For each request $req \in Q$, select the least-loaded instance $p_{target}$ from the candidate list. If $U_{p_{target}} < \delta_L$, assign $req$ to $p_{target}$; otherwise, select the instance with the smallest queue length and assign it to that instance. After assignment, increment $load[p_{target}]$ by the estimated load contribution of $req$.
    \item[4] \textbf{Finalize plan} (lines 21–23): Generate the dispatch map $D^*$ using \texttt{GetDispatchMap} and return it to the orchestration component for execution.
\end{enumerate}

The computational complexity per scheduling cycle is
\begin{equation}
O(|P| \log |P| + |Q|),
\end{equation}
where $|P|$ is the number of prefill instances and $|Q|$ is the number of requests in the queue. The $O(|P|\log|P|)$ term results from sorting instances by current load and queue length, and the $O(|Q|)$ term covers the per-request assignment loop. In typical deployments, $|P|$ is small (dozens of GPUs), making the algorithm suitable for real-time scheduling.

By removing cache hit rate from the routing criteria through the Global KV Cache Store, this scheduler focuses solely on balancing load and minimizing queuing delay. This design reduces complexity, avoids hotspot formation in prefill nodes, and ensures stable throughput under dynamic workloads.

% \input{algorithm1}

% \input{algorithm2}
% Based on our performance models and experimental observations, we design an adaptive migration algorithm that dynamically adjusts module allocation based on real-time system metrics. The algorithm operates on a feedback control loop that monitors resource utilization, request queue lengths, and performance metrics to make migration decisions.

% The algorithm maintains a target utilization balance between prefill and decode instances while minimizing migration frequency to reduce overhead. Key algorithmic components include:

% \textbf{Utilization Monitoring}: Continuous tracking of CPU, GPU, and memory utilization across all instances.

% \textbf{Load Prediction}: Short-term forecasting of request patterns based on historical data.

% \textbf{Migration Decision}: Cost-benefit analysis comparing current performance against potential improvements from module reallocation.

% \textbf{Migration Execution}: Coordinated transfer of modules with minimal service disruption.

% The algorithm employs hysteresis mechanisms to prevent oscillatory behavior and includes safety constraints to ensure system stability during migration operations. Performance thresholds trigger migration decisions, with different sensitivity levels for upward and downward scaling to balance responsiveness with stability.

\section{Evaluations}\label{sec 5}

In this section, we present a comprehensive empirical evaluation of the proposed \textbf{BanaServe} framework. Our study compares BanaServe against two state-of-the-art LLM serving systems across multiple model architectures, context length scenarios, and workload intensities. We detail the experimental setup, benchmark selection, evaluation metrics, and load testing methodology, followed by an in-depth analysis of performance results including throughput, end-to-end latency, and scalability.

\subsection{Experimental Setup}

To comprehensively evaluate the effectiveness of the BanaServe framework, we conducted an extensive empirical study comparing its performance against two leading state-of-the-art inference systems: DistServe~\cite{zhong2024distserve} and vLLM~\cite{vllm}. Our evaluation methodology employs a multi-dimensional performance analysis across diverse operational conditions, model architectures, and workload characteristics.

\subsubsection{Model Selection and Configurations}

We strategically selected two representative large language models to ensure coverage of different architectural paradigms at the same parameter scale. Table~\ref{tab:model_configs} summarizes their key characteristics. The LLaMA-13B\footnote{\url{https://huggingface.co/meta-llama/Llama-2-13b}} model serves as the primary evaluation target within the LLaMA family, enabling us to assess how BanaServe handles large-scale decoder-only transformers with increased model complexity. Additionally, we include OPT‑13B\footnote{\url{https://huggingface.co/facebook/opt-13b}} to facilitate cross-architecture validation. Although both models have the same parameter count, OPT‑13B employs alternative architectural optimizations and training methodologies, allowing us to examine the feasibility of BanaServe across distinct transformer implementations.
\begin{table}[htbp]
\centering
\caption{Model configurations used in experimental evaluation}
\label{tab:model_configs}
\begin{tabular}{lccc}
\toprule
\textbf{Model} & \textbf{Parameters} & \textbf{Architecture} & \textbf{Evaluation Purpose} \\
\midrule
LLaMA-13B & 13B & Decoder-only & Intra-family performance evaluation \\
OPT-13B   & 13B & Decoder-only & Cross-architecture validation       \\
\bottomrule
\end{tabular}
\end{table}
\vspace{-0.5cm}
% 这里把aplaca和 longbench的图加进去
\subsubsection{Benchmark Selection and Workload Scenarios}

% Our experimental design leverages two industry-standard benchmarks that comprehensively evaluate different aspects of inference performance under varying context lengths. These benchmarks represent complementary workload patterns commonly encountered in production deployments.
Our experimental design leverages two publicly available, industry-standard benchmarks: \textbf{Alpaca}\footnote{\url{https://github.com/tatsu-lab/stanford_alpaca}} and \textbf{LongBench}\footnote{\url{https://github.com/THUDM/LongBench}}, which together comprehensively evaluate inference performance under varying context lengths. These benchmarks represent complementary workload patterns commonly encountered in production deployments.

\begin{figure}[htbp]
    \centering
    \begin{subfigure}[b]{0.45\textwidth}
        \centering
        \includegraphics[width=\textwidth]{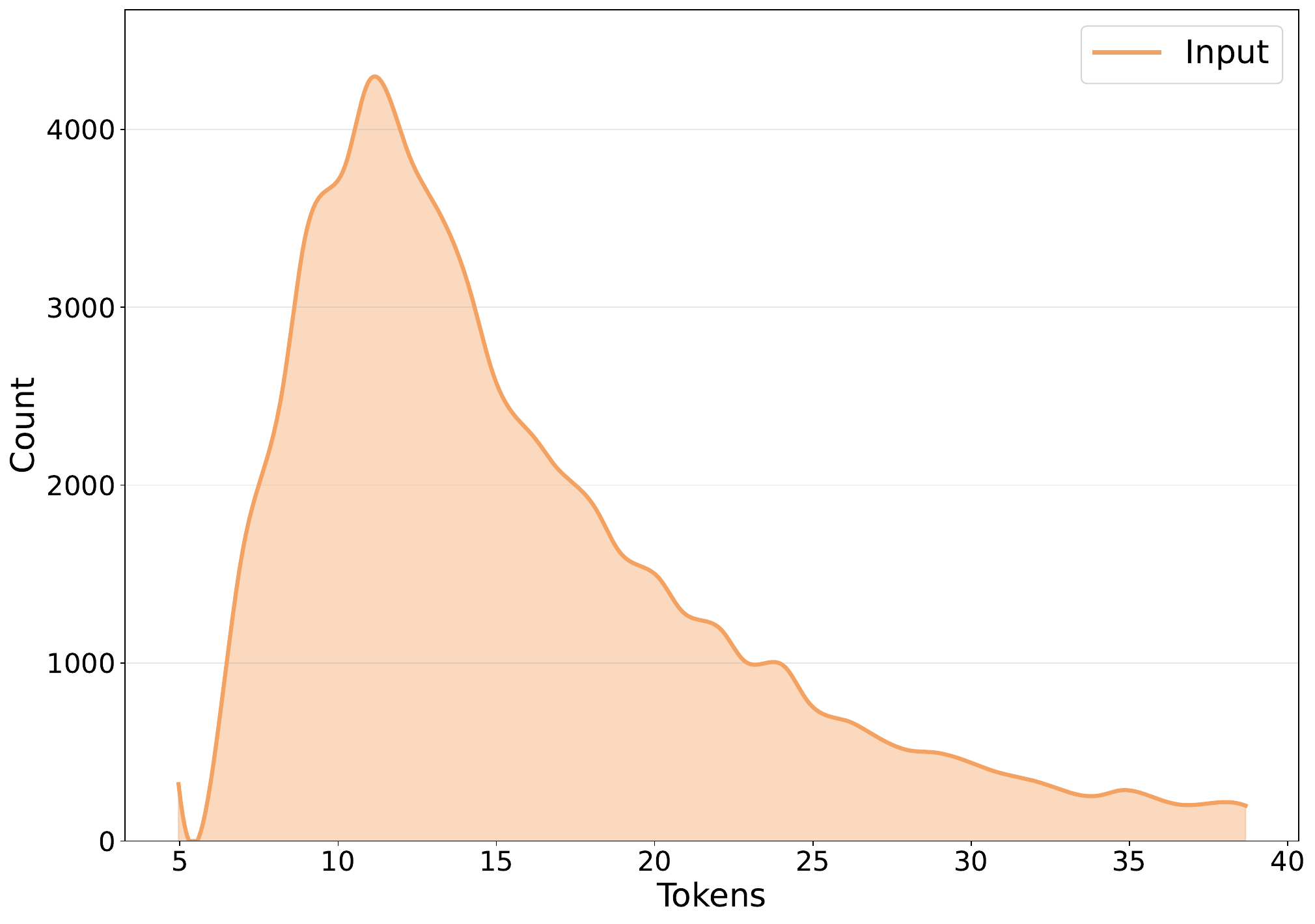}
        \caption{Alpaca}
        \label{fig:Alpaca}
    \end{subfigure}
    \hfill
    \begin{subfigure}[b]{0.45\textwidth}
        \centering
        \includegraphics[width=\textwidth]{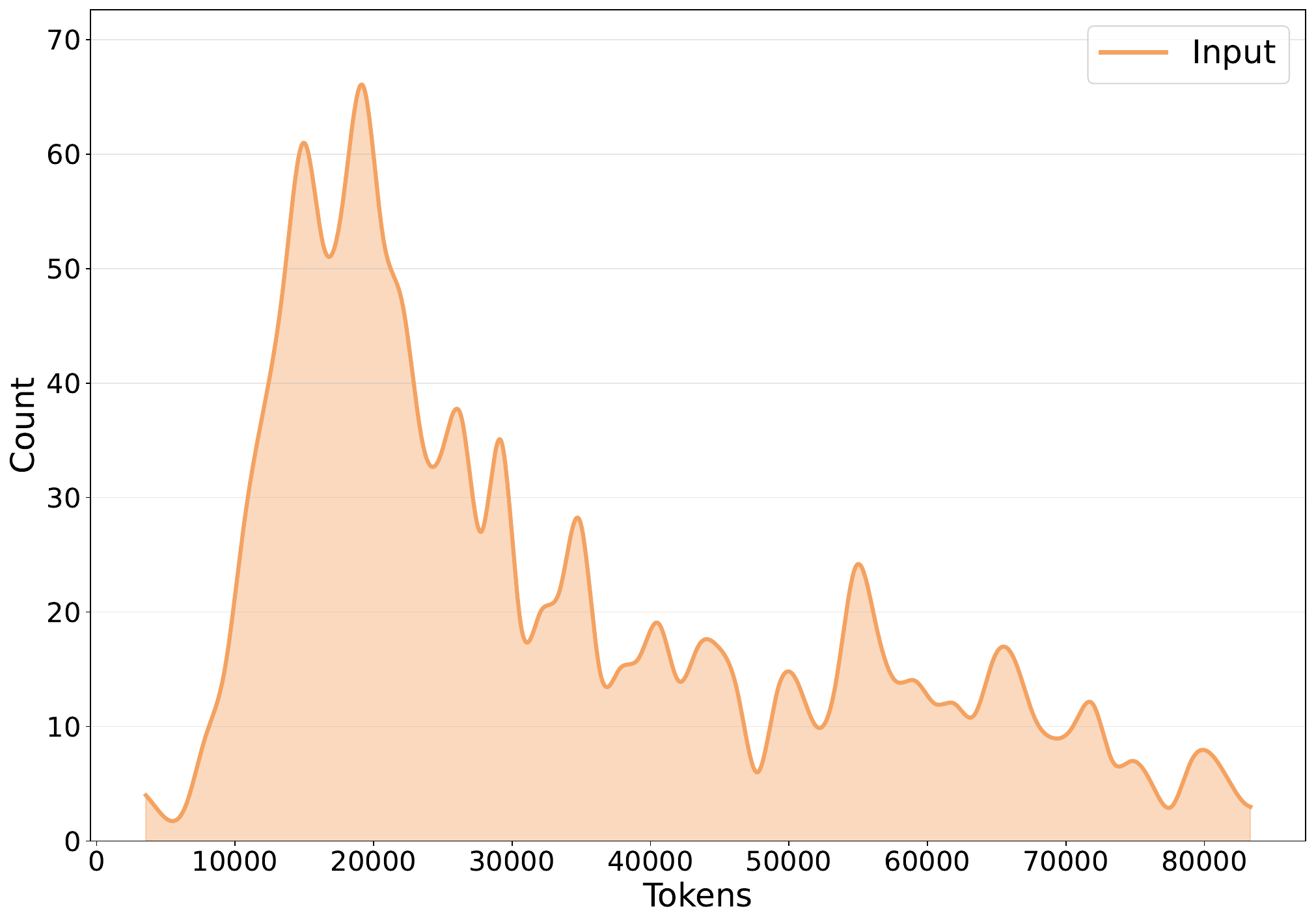}
        \caption{LongBench}
        \label{fig:LongBench}
    \end{subfigure}

    \caption{
    Input length distributions for the two benchmarks used in our evaluation. 
    (\subref{fig:Alpaca}) The Alpaca dataset consists primarily of short-context instruction-following inputs, with sequence lengths ranging from 5 to 40 tokens in our experiments.  
    (\subref{fig:LongBench}) The LongBench dataset contains long-context inputs with sequence lengths spanning from roughly 2{,}000 to over 85{,}000 tokens, covering diverse multi-document and multi-task scenarios.  
    In all experiments, the maximum \emph{output length} is capped at 512 tokens to ensure consistency across benchmarks, avoid excessive generation latency, and maintain comparability of throughput and latency metrics while focusing on input-length variability.
    }
    \label{fig:length}
\end{figure}

For \textbf{short-context} evaluation, we employ the Alpaca benchmark, which consists of 52{,}000 instruction-following examples generated using self-instruct techniques from Stanford. Alpaca represents typical production inference workloads with input sequences ranging from 4 to 50 tokens as shown in Figure~\ref{fig:Alpaca}, capturing the characteristics of interactive applications such as chatbots, code assistants, and Q\&A systems. The benchmark's diverse instruction types, spanning classification, generation, editing, and reasoning tasks, provide a broad assessment of inference efficiency across different computational patterns. These short-context experiments particularly stress the system's request scheduling mechanisms, time-to-first-token optimization, and serving infrastructure overhead under varied load conditions.

For \textbf{long-context} evaluation, we utilize LongBench, a comprehensive multi-task benchmark specifically designed to assess long-context understanding capabilities of large language models. LongBench encompasses 21 diverse tasks across 6 categories, with context lengths from \(\sim\)2{,}000 to over 85{,}000 tokens, as shown in Figure~\ref{fig:LongBench}. The benchmark includes multi-document QA (HotpotQA, 2WikiMultihopQA), single-document QA (NarrativeQA, Qasper), summarization (GovReport, MultiNews), few-shot learning tasks, synthetic tasks probing specific long-context abilities, and code completion tasks. This variation in sequence length acts as a stress test for KV Cache management and memory optimization strategies. Its multilingual nature, including English, Chinese, and programming languages, ensures our evaluation captures performance variations across different tokenization schemes and vocabulary distributions.

We adopt a comprehensive metric suite to capture different aspects of system performance. Throughput, measured in tokens per second, quantifies the raw processing capability of each framework and directly reflects the efficiency of GPU utilization and batch processing strategies. This metric is particularly important for understanding system capacity under high-load scenarios. Total time encompasses the end-to-end latency from request submission to complete response generation, capturing all system overheads including model loading, attention computation, KV Cache management, and result post-processing. For user-facing applications, we measure average latency decomposed into TTFT and inter-token latency, providing critical insights into perceived responsiveness. TTFT is especially crucial for interactive applications where users expect immediate feedback, while inter-token latency determines the smoothness of the generation process.

\subsubsection{Load Testing Methodology}

To evaluate system behavior under diverse operational conditions, we implemented a progressive load testing strategy with request rates systematically varied from 1 to 20 requests per second (RPS). This range captures system behavior from under-utilized states where latency is primarily determined by computational efficiency, through moderate loads where scheduling and batching strategies become critical, to fully saturated states where queueing delays and resource contention dominate performance. We employed a Poisson arrival process to simulate realistic traffic patterns, as this better represents the bursty nature of real-world inference requests compared to uniform arrivals, allowing us to evaluate system resilience under variable load conditions.

Each experimental configuration underwent rigorous testing with a 60-second warm-up period to ensure system stability, populate all caches, and reach steady-state performance before measurement collection. To ensure statistical validity, all experiments were repeated five times with different random seeds, and we report mean values along with 95\% confidence intervals. This methodology allows us to distinguish genuine performance differences from random variations in system behavior and ensures reproducibility of our results.

\subsection{Results and Analysis}

In this section, we present a comprehensive evaluation of BanaServe's end-to-end performance against the baseline systems across different workload scenarios and model architectures.

\subsubsection{Baselines}
We compare BanaServe with two state-of-the-art LLM serving systems:

\textbf{vLLM~\cite{vllm}:} A widely adopted serving system in both academia and industry that employs continuous batching and PagedAttention for memory-efficient KV Cache management. However, vLLM's monolithic architecture struggles to balance the competing demands of prefill and decode phases, particularly under stringent latency requirements.

\textbf{DistServe~\cite{zhong2024distserve}:} A disaggregated serving system that separates prefill and decode computations into specialized instances. While this approach mitigates interference between phases, it introduces additional communication overhead and requires careful coordination between disaggregated components.

% Figure 2: short-context Experiments
% Figure: short-context Experiments - LLaMA
\begin{figure}[htbp]
    \centering
    
    \begin{subfigure}[b]{0.32\textwidth}
        \centering
        \includegraphics[width=\textwidth]{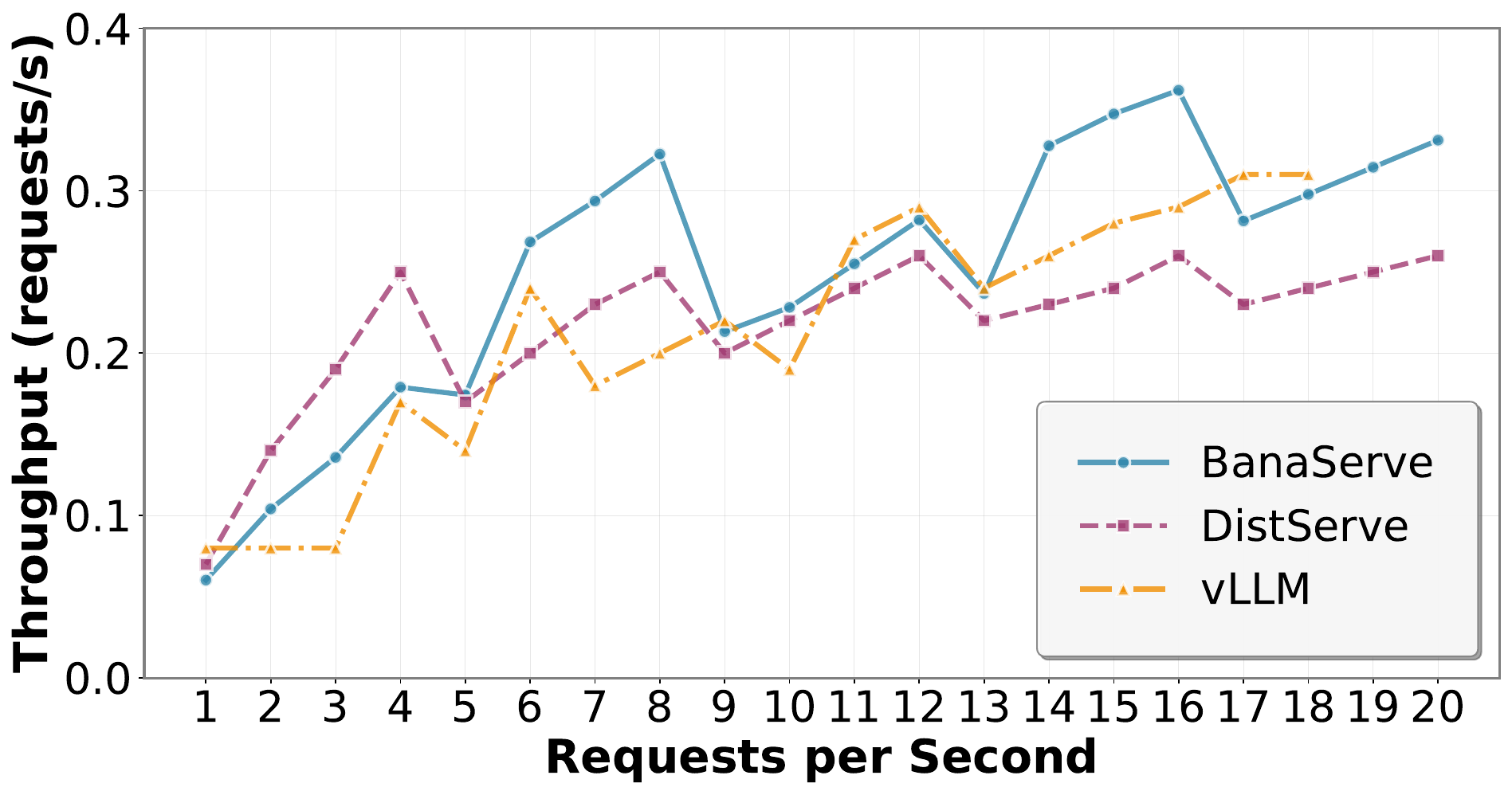}
        \caption{Throughput}
        \label{fig:llama_short_throughput}
    \end{subfigure}
    \hfill
    \begin{subfigure}[b]{0.32\textwidth}
        \centering
        \includegraphics[width=\textwidth]{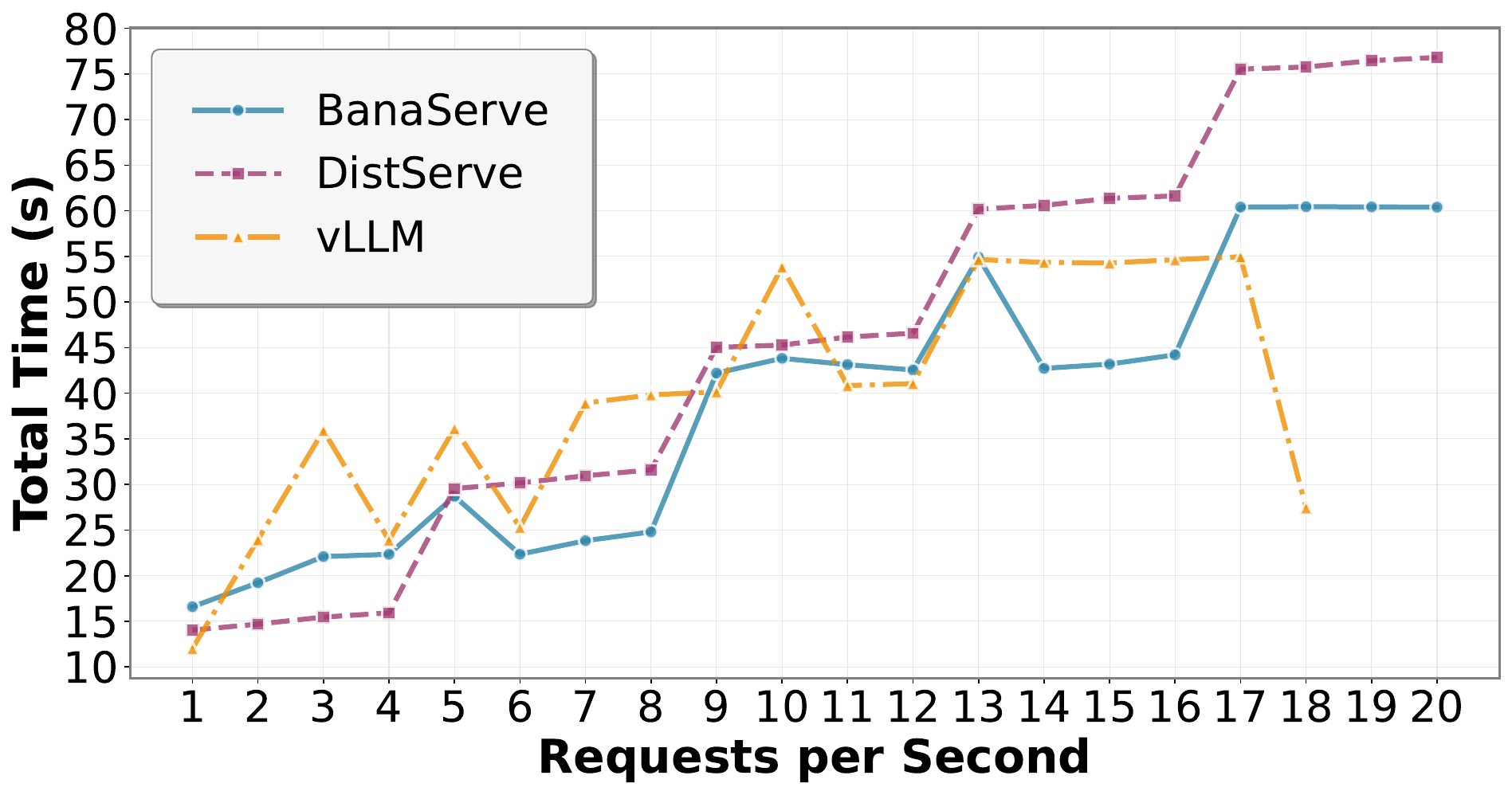}
        \caption{Total Time}
        \label{fig:llama_short_total_time}
    \end{subfigure}
    \hfill
    \begin{subfigure}[b]{0.32\textwidth}
        \centering
        \includegraphics[width=\textwidth]{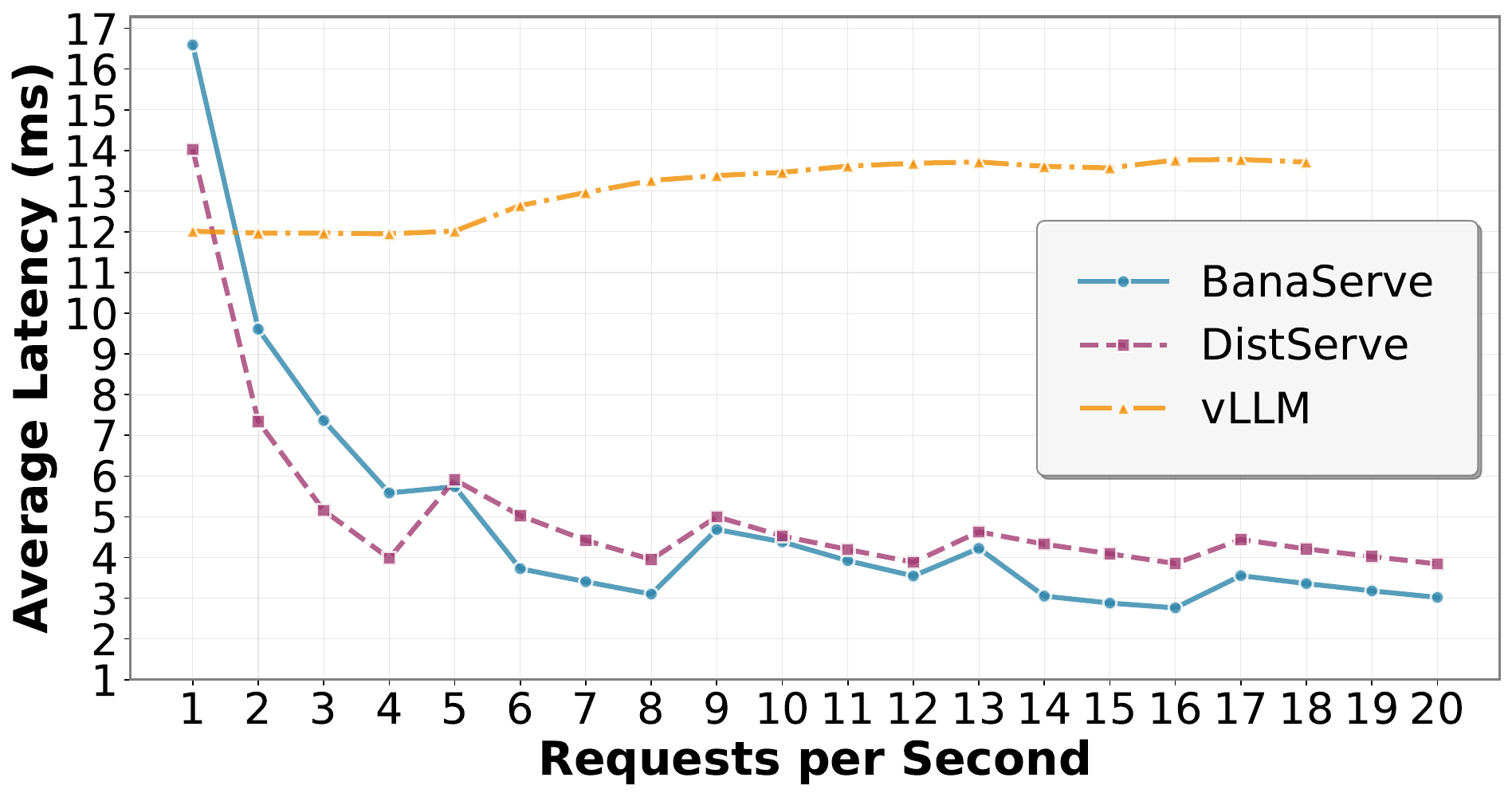}
        \caption{Average Latency}
        \label{fig:llama_short_avg_latency}
    \end{subfigure}
    
    \caption{Short-context performance results for \textbf{LLaMA-13B}. 
    Experiments compare three inference frameworks: BanaServe, DistServe, and vLLM across different RPS (requests per second) loads.}
    \label{fig:short_context_llama}
\end{figure}

% Figure: short-context Experiments - OPT
\begin{figure}[htbp]
    \centering
    
    \begin{subfigure}[b]{0.32\textwidth}
        \centering
        \includegraphics[width=\textwidth]{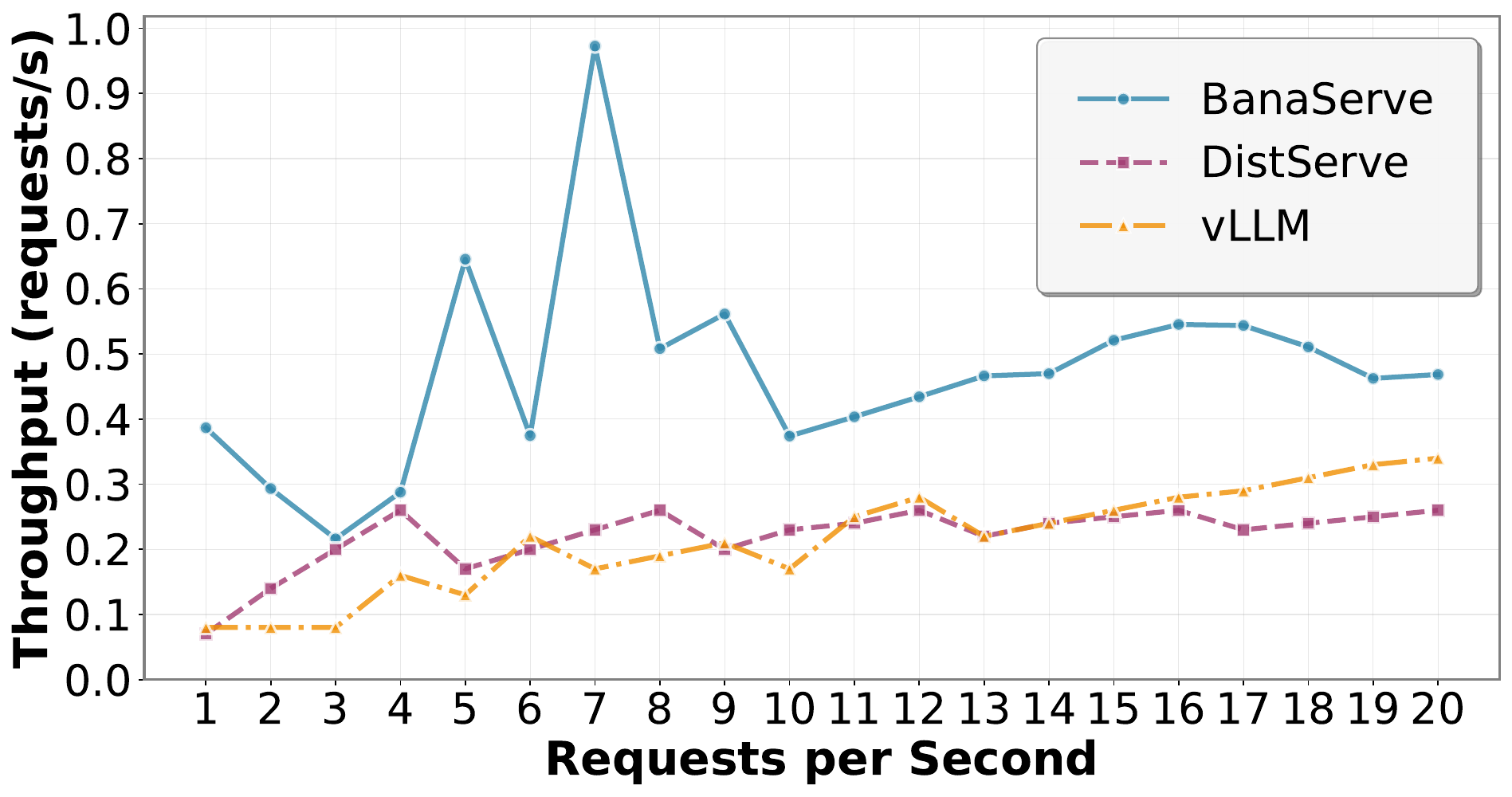}
        \caption{Throughput}
        \label{fig:opt_short_throughput}
    \end{subfigure}
    \hfill
    \begin{subfigure}[b]{0.32\textwidth}
        \centering
        \includegraphics[width=\textwidth]{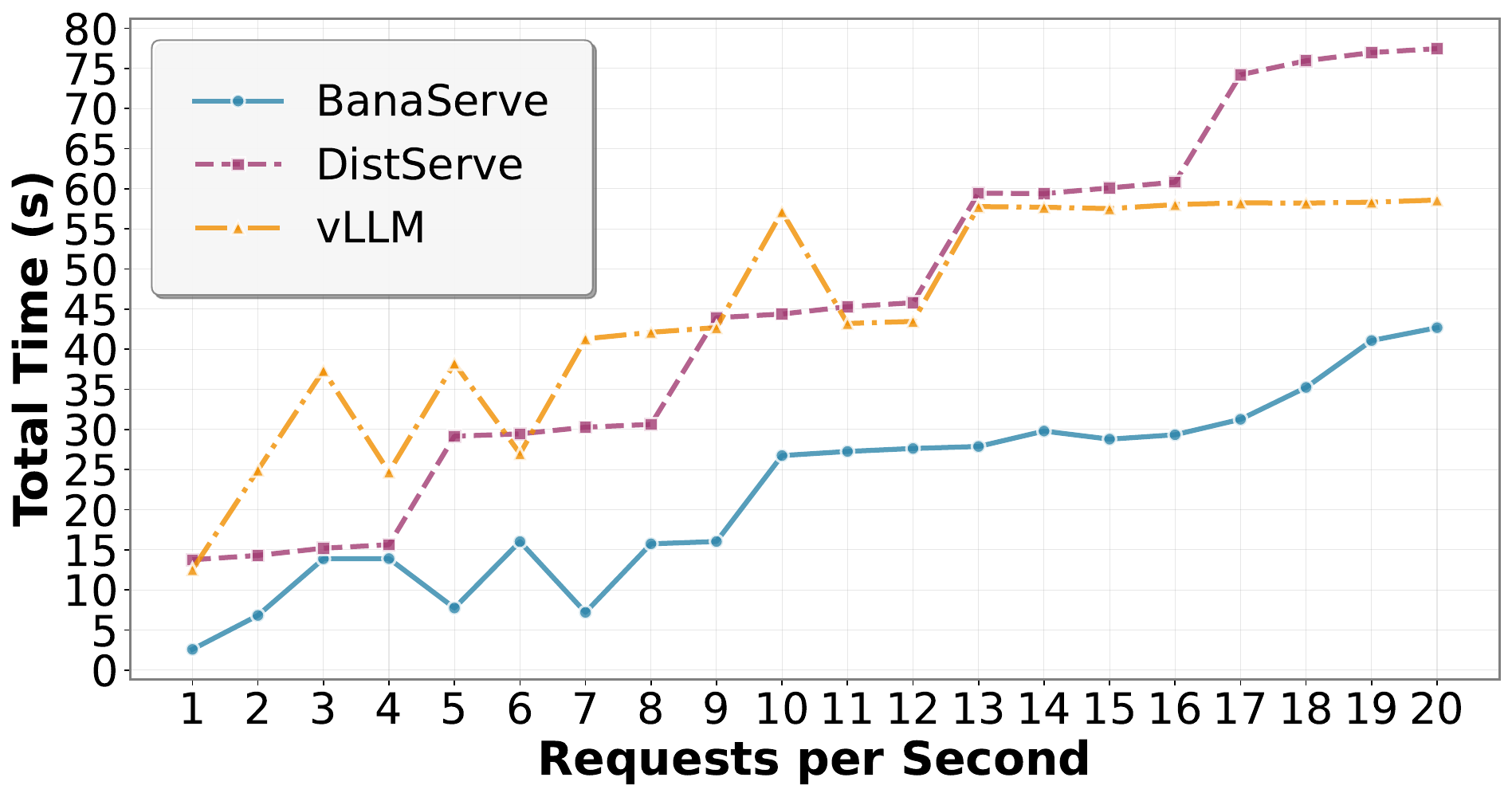}
        \caption{Total Time}
        \label{fig:opt_short_total_time}
    \end{subfigure}
    \hfill
    \begin{subfigure}[b]{0.32\textwidth}
        \centering
        \includegraphics[width=\textwidth]{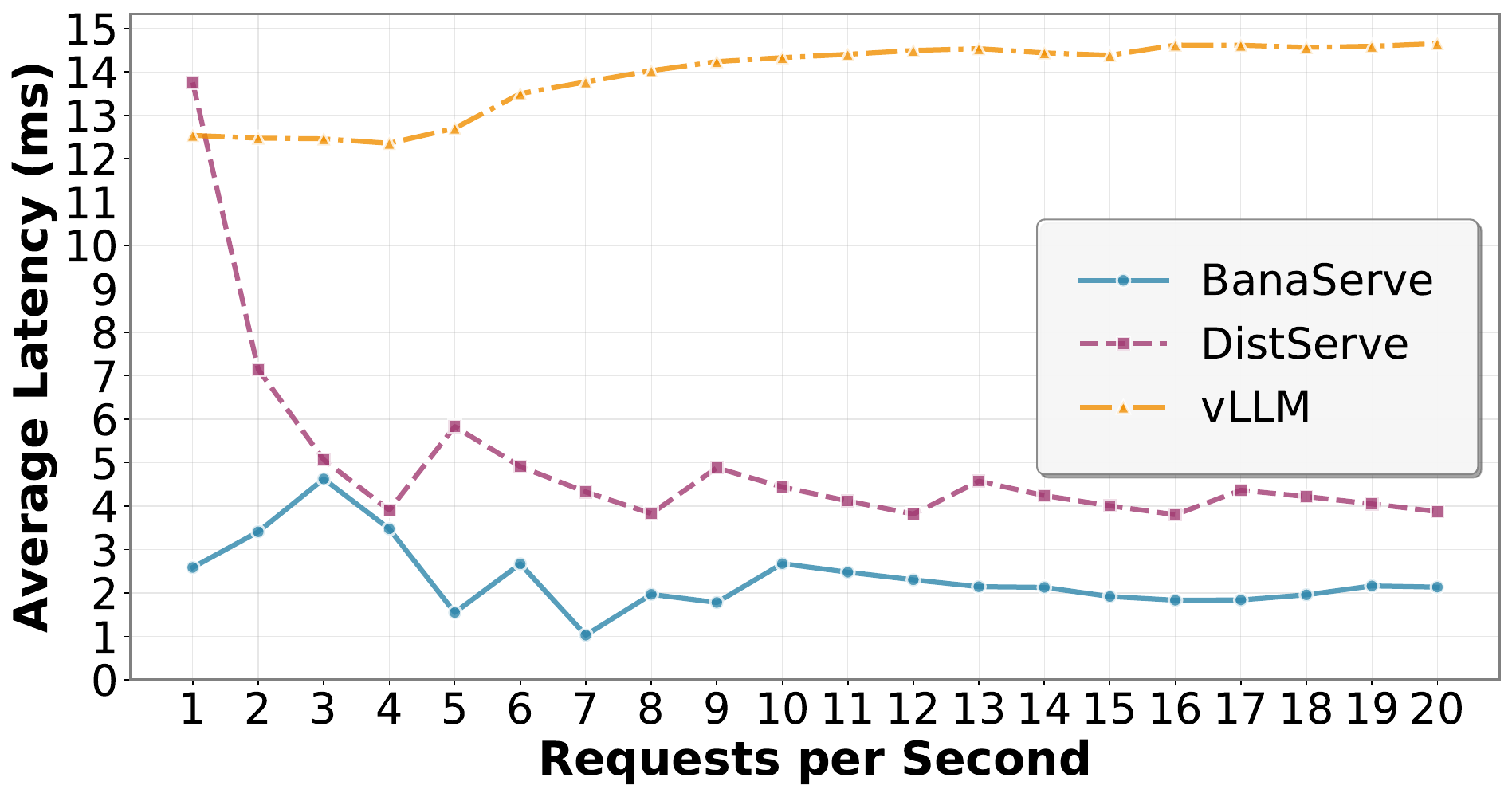}
        \caption{Average Latency}
        \label{fig:opt_short_avg_latency}
    \end{subfigure}
    
    \caption{Short-context performance results for \textbf{OPT-13B}.
    Experiments compare three inference frameworks: BanaServe, DistServe, and vLLM across different RPS loads.}
    \label{fig:short_context_opt}
\end{figure}

\subsubsection{Performance}

We evaluate \textbf{BanaServe} across two representative LLM architectures (LLaMA-13B and OPT-13B) and two context regimes 
(LongBench for extended sequences, Alpaca for short production-style workloads). 
Performance is reported across varying input rates (1–20 RPS) and expressed in terms of three primary metrics: 
throughput (tokens/s), total processing time, and average per-request latency.

\textbf{LLaMA-13B with Short-context.} 
Figure~\ref{fig:short_context_llama} presents results on the Alpaca benchmark with short, production-style inputs. 
Across the entire RPS spectrum, BanaServe delivers consistently higher throughput, ranging from \(1.1{\times}\) to \(1.2{\times}\) compared to both DistServe and vLLM, while maintaining lower latency. 
Reducing per-request latency ensures high responsiveness in interactive settings, and total processing times remain lower under all loads. 
These improvements stem from BanaServe’s minimized scheduling overhead and efficient KV Cache handling, which allow rapid context switches without incurring significant pipeline stalls.

\textbf{OPT-13B with Short-context.} 
For OPT-13B under the same short-context workload as shown in Figure~\ref{fig:short_context_opt}, performance gains are markedly higher. 
BanaServe achieves throughput improvements ranging from \(2.8{\times}\) to \(3.9{\times}\) compared to DistServe, and up to \(3.9{\times}\) compared to vLLM. 
Average latency reductions are likewise substantial, with decreases of \(3.9\%\) to \(78.4\%\) relative to vLLM, and \(1.4\%\) to \(70.1\%\) relative to DistServe. 
These results highlight BanaServe’s efficiency in high-frequency context-switch scenarios, in which its batching and cache reuse strategies enable near-saturation of GPU utilization while maintaining responsiveness.

% Figure: Long-context Experiments - LLaMA
\begin{figure}[htbp]
    \centering
    
    \begin{subfigure}[b]{0.32\textwidth}
        \centering
        \includegraphics[width=\textwidth]{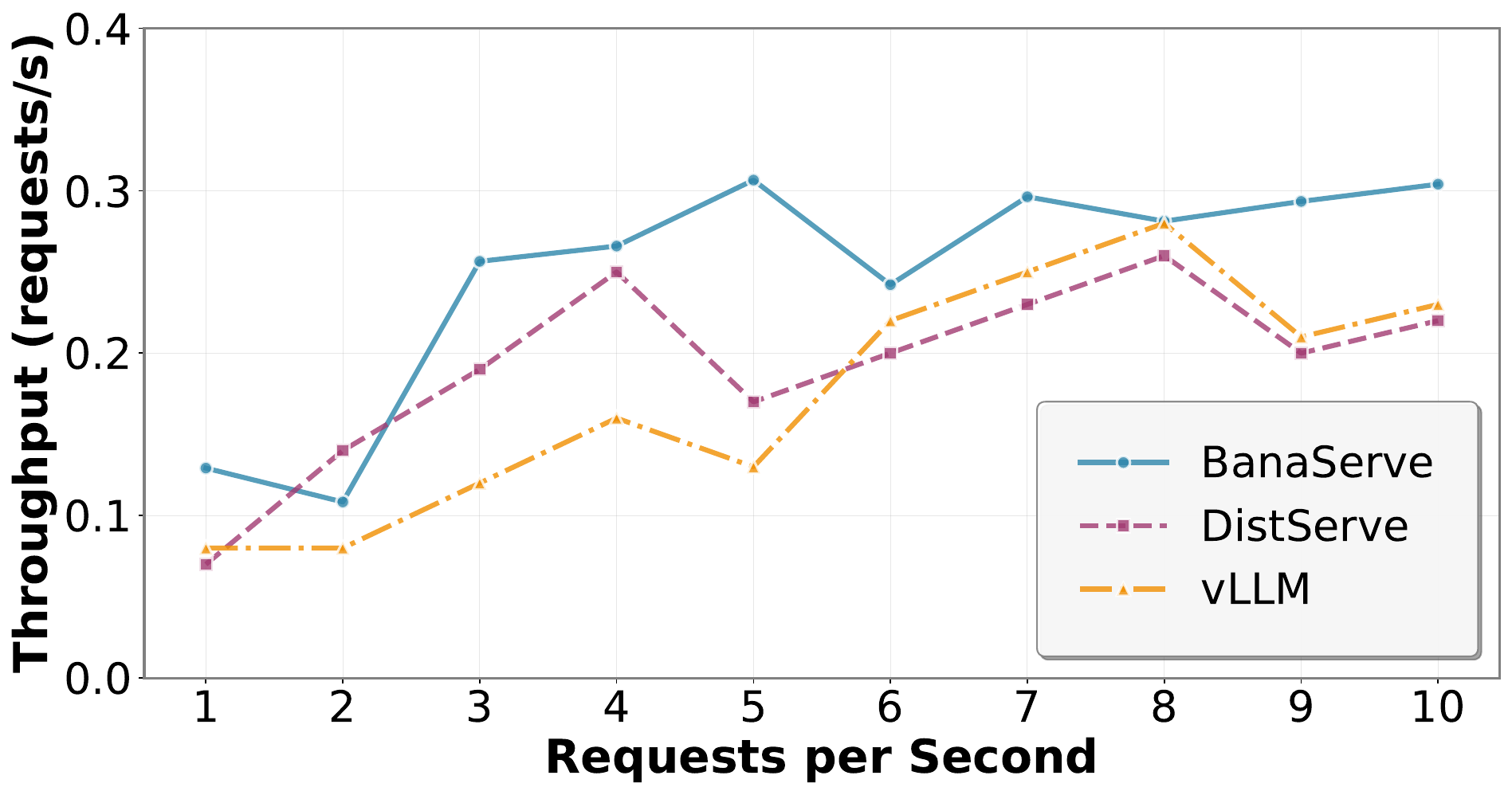}
        \caption{Throughput}
        \label{fig:llama_long_throughput}
    \end{subfigure}
    \hfill
    \begin{subfigure}[b]{0.32\textwidth}
        \centering
        \includegraphics[width=\textwidth]{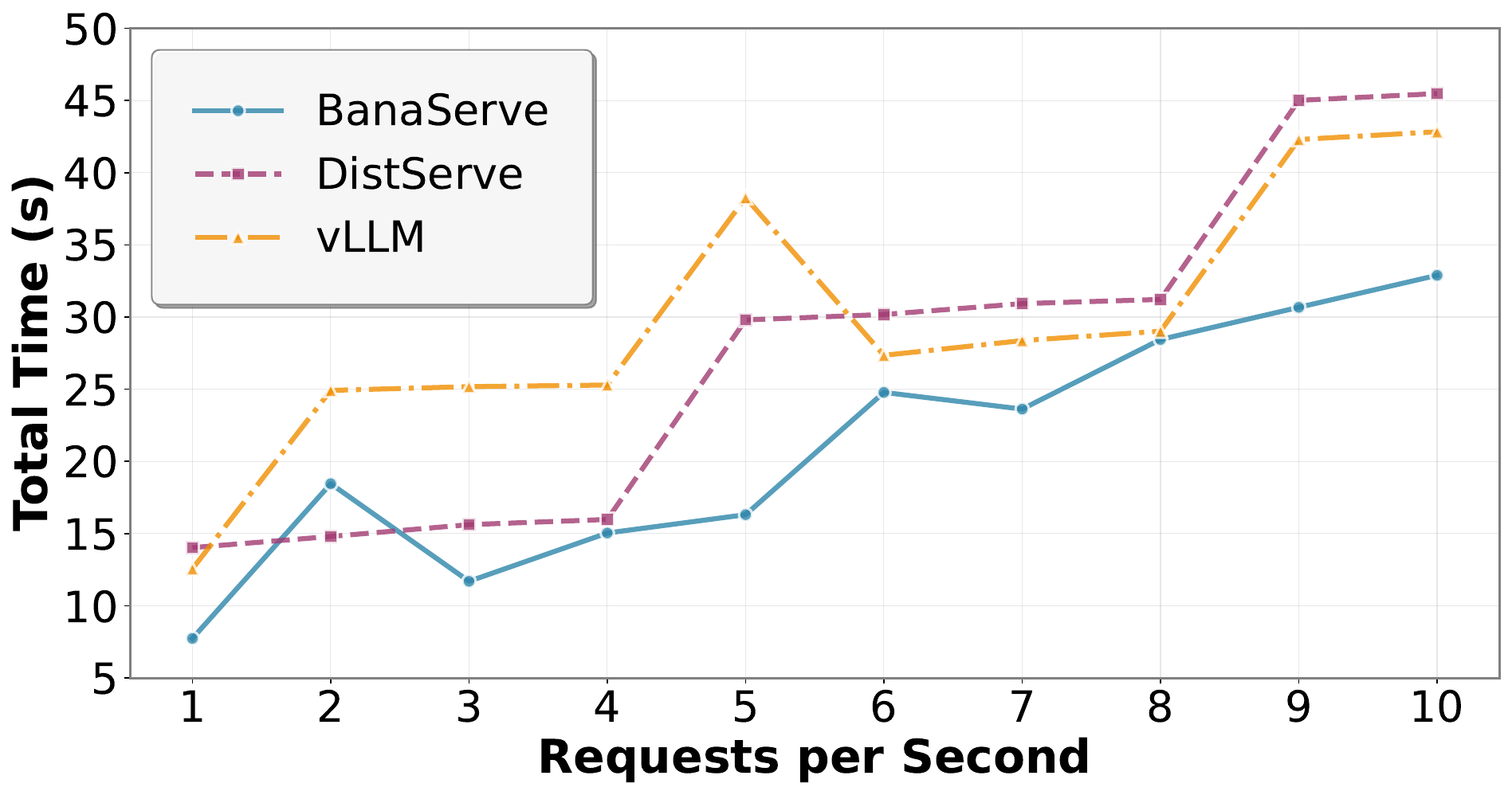}
        \caption{Total Time}
        \label{fig:llama_long_total_time}
    \end{subfigure}
    \hfill
    \begin{subfigure}[b]{0.32\textwidth}
        \centering
        \includegraphics[width=\textwidth]{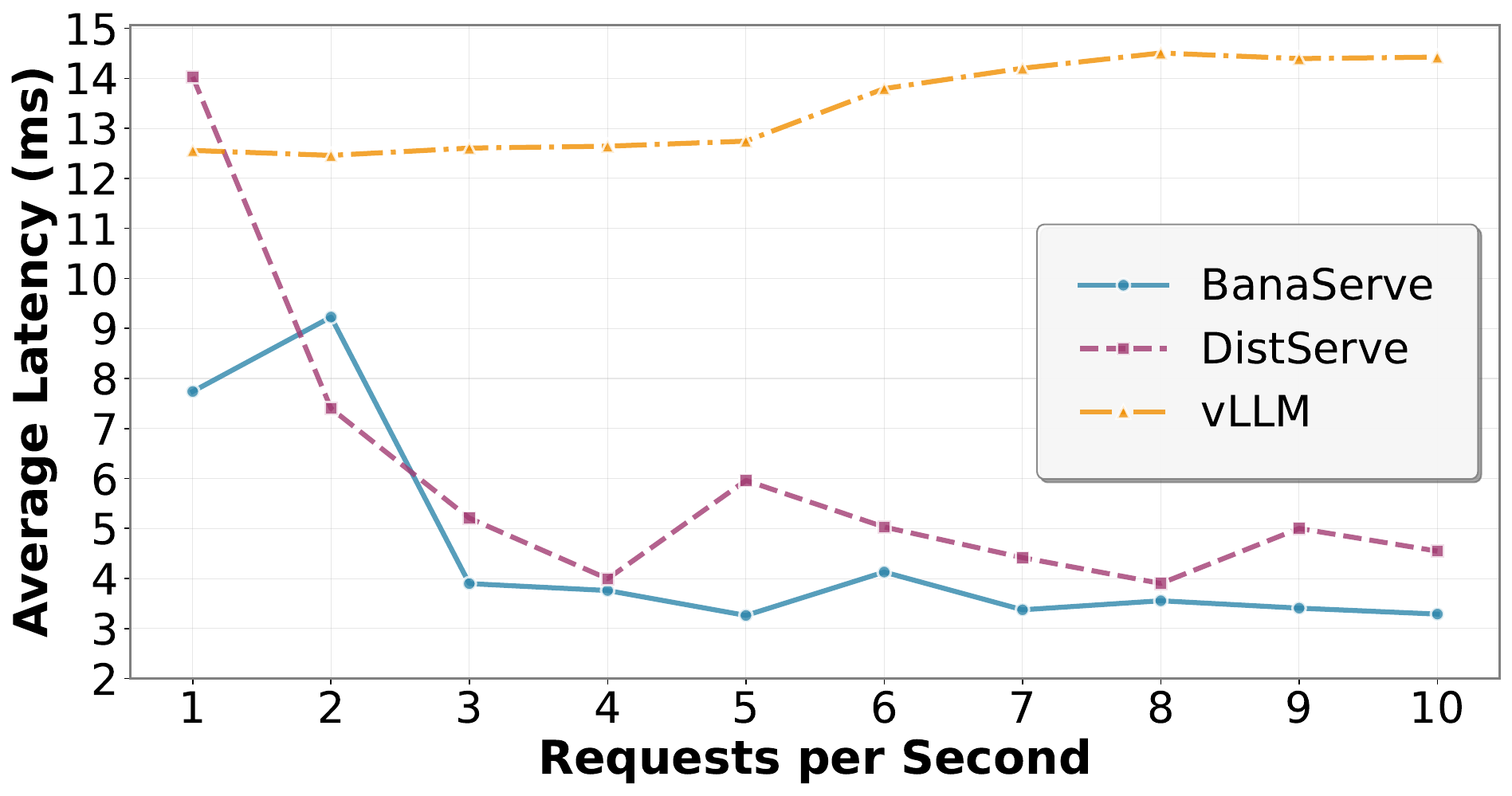}
        \caption{Average Latency}
        \label{fig:llama_long_avg_latency}
    \end{subfigure}
    
    \caption{Long-context performance results for \textbf{LLaMA-13B}.
    Experiments compare three inference frameworks: BanaServe, DistServe, and vLLM across different RPS (requests per second) loads.}
    \label{fig:long_context_llama}
\end{figure}

% Figure: Long-context Experiments - OPT
\begin{figure}[htbp]
    \centering
    
    \begin{subfigure}[b]{0.32\textwidth}
        \centering
        \includegraphics[width=\textwidth]{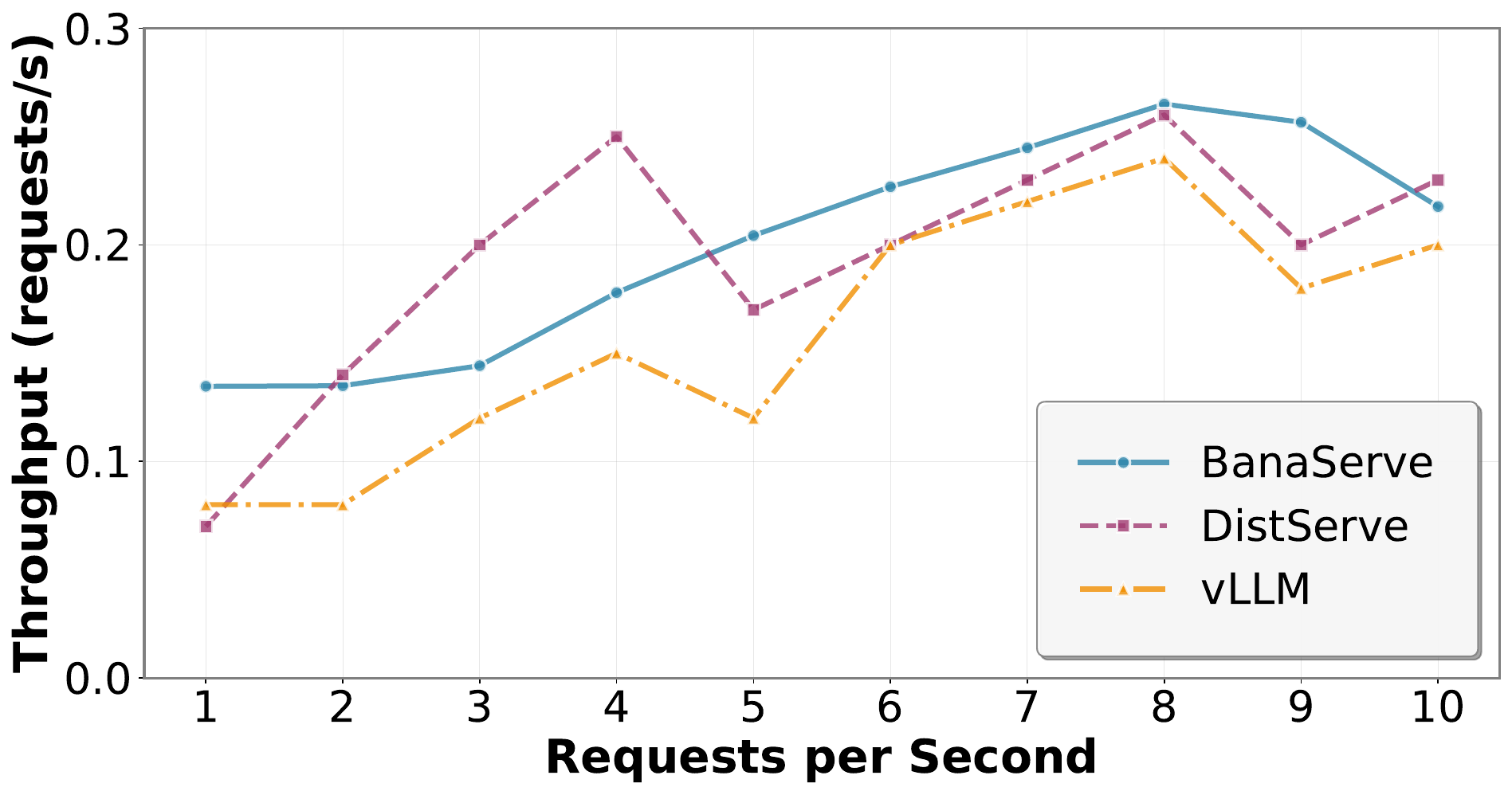}
        \caption{Throughput}
        \label{fig:opt_long_throughput}
    \end{subfigure}
    \hfill
    \begin{subfigure}[b]{0.32\textwidth}
        \centering
        \includegraphics[width=\textwidth]{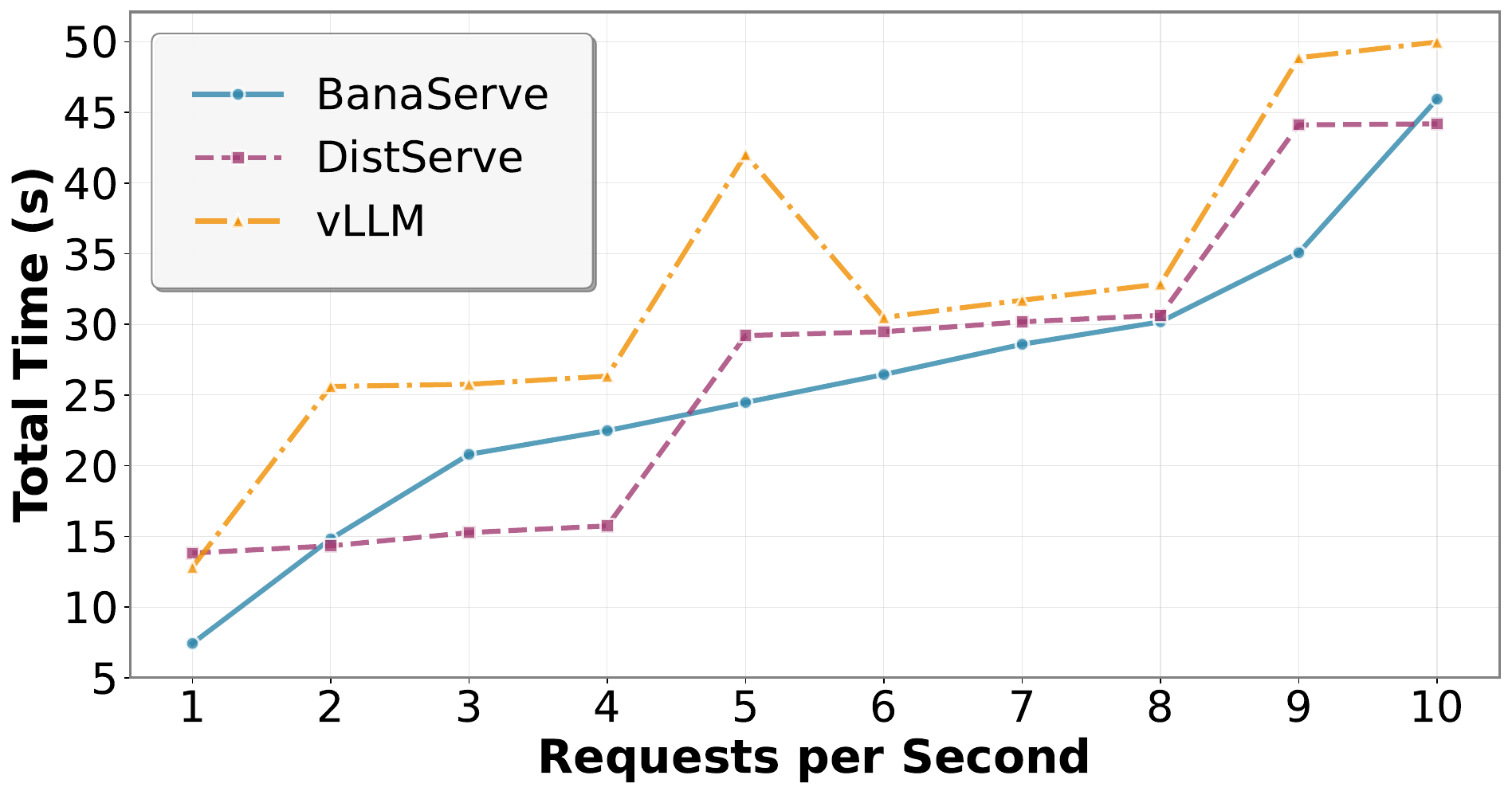}
        \caption{Total Time}
        \label{fig:opt_long_total_time}
    \end{subfigure}
    \hfill
    \begin{subfigure}[b]{0.32\textwidth}
        \centering
        \includegraphics[width=\textwidth]{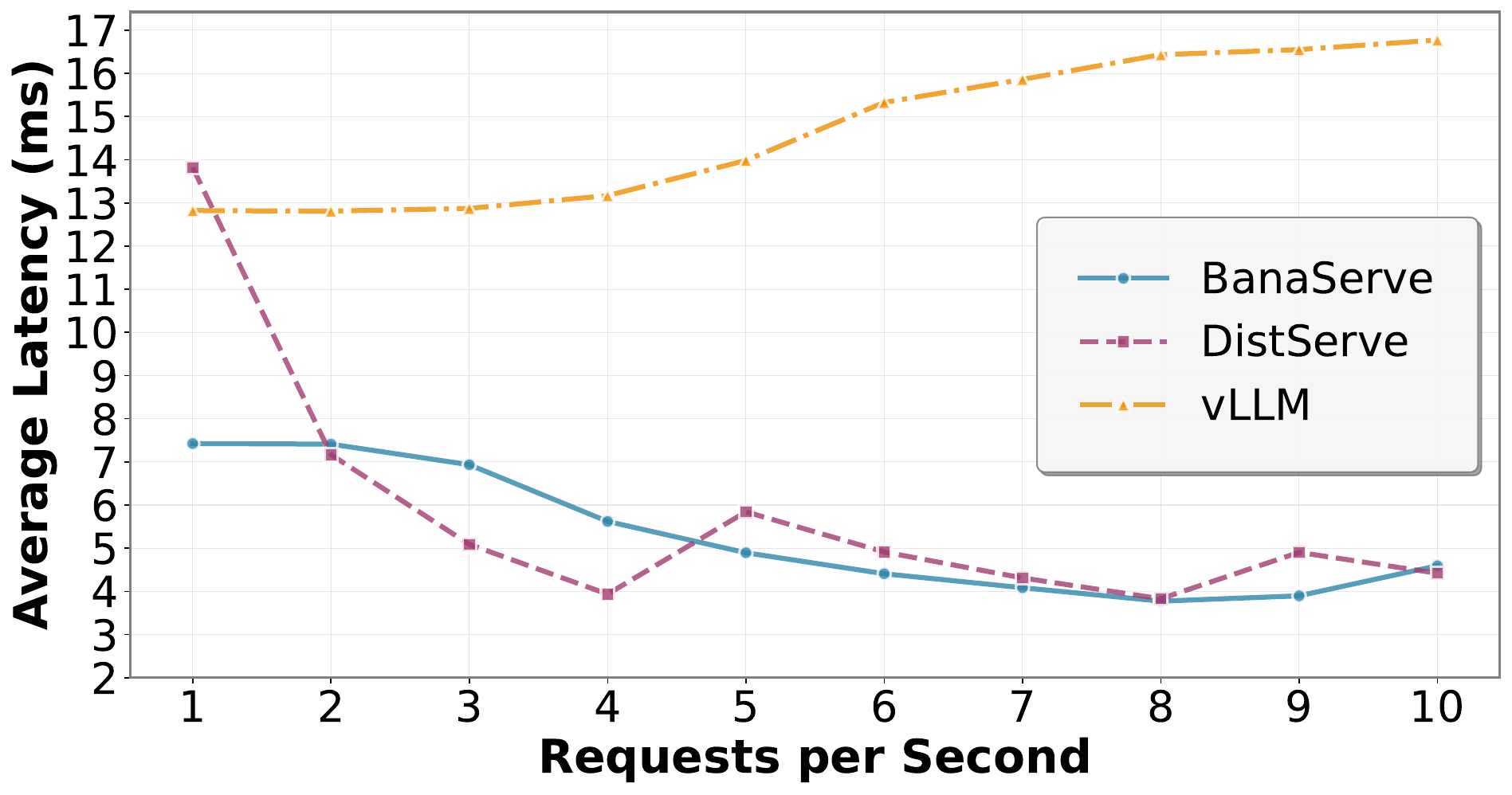}
        \caption{Average Latency}
        \label{fig:opt_long_avg_latency}
    \end{subfigure}
    
    \caption{Long-context performance results for \textbf{OPT-13B}.
    Experiments compare three inference frameworks: BanaServe, DistServe, and vLLM across different RPS loads.}
    \label{fig:long_context_opt}
\end{figure}

\textbf{LLaMA-13B with Long-context.} 
On the LongBench benchmark with extended sequences as demonstrated in Figure~\ref{fig:long_context_llama}, BanaServe delivers throughput improvements of \(1.3{\times}\) to \(1.5{\times}\) over DistServe and up to \(1.5{\times}\) over vLLM. 
Latency reductions range from \(20.6\%\) to \(65.3\%\) compared to vLLM and \(1.4\%\) to \(20.6\%\) compared to DistServe. 
The performance gap becomes more pronounced under high load conditions (10–20 RPS), where conventional systems suffer from cache contention and pipeline blocking, whereas BanaServe’s unified architecture reduces inter-component communication overhead and exploits dynamic batching to balance computational and memory resources across phases.

\textbf{OPT-13B with Long-context.} 
In the same LongBench setting as illusrated in Figure~\ref{fig:long_context_opt}, BanaServe sustains throughput increases of \(1.1{\times}\) to \(1.3{\times}\) relative to DistServe and up to \(1.3{\times}\) relative to vLLM. 
Although these gains are smaller than those observed for LLaMA-13B, they remain consistent across different load levels and can be attributed to reduced KV Cache fetch latency as well as more efficient memory allocation patterns. 
The latency improvements follow the same pattern, ensuring stable quality of service even when processing large context windows.

\textbf{Scalability Analysis.} As request rates increase from 1 to 20 RPS, BanaServe maintains consistent performance advantages across all configurations. Under light loads (1-5 RPS), the performance gap is primarily attributed to BanaServe's efficient scheduling and reduced system overhead. As load increases (10-20 RPS), BanaServe's superior batching strategies and optimized memory management enable it to sustain higher throughput while other systems experience performance degradation due to resource contention and inefficient cache utilization.

The results validate that BanaServe's unified yet flexible architecture successfully combines the benefits of monolithic systems (low overhead, efficient resource sharing) with those of disaggregated systems (phase-specific optimization, reduced interference), achieving superior performance across diverse workload characteristics and operational conditions.

\section{Conclusions, Discussions and Future Work}\label{sec:conclusion}
This paper presents \textbf{BanaServe}, a inference serving framework tailored to address compute and memory imbalance and KV Cache management challenges in PD disaggregated architectures for LLMs. We introduce two key components: a Global KV Cache Store for cross-instance KV Cache sharing, and two adaptive runtime algorithms, Dynamic Migration and Load-aware Request Scheduling, that coordinate PD execution without reliance on prefix cache hit rate. Our design leverages layer-wise pipeline overlap to hide communication latencies, granularity aware migration to balance workloads at both coarse (layer-level) and fine (attention-level) levels, and load-oriented scheduling to distribute requests across heterogeneous GPU clusters. Validation using the Alpaca and LongBench benchmarks demonstrates significant reduction in first-token latency, improved throughput under dynamic workloads, and robust performance across short- and long-context inference scenarios. These results confirm the practicality and scalability of BanaServe in production-like serving environments.

Although BanaServe demonstrates strong potential for efficient LLM inference serving in disaggregated architectures, its current implementation still faces several limitations. First, the system’s scheduling and migration policies lack comprehensive optimization for heterogeneous hardware, such as mixed-precision accelerators and specialized communication fabrics, which may result in underutilized cluster resources. Second, its workload management is primarily reactive, relying on instantaneous load metrics without predictive modeling from historical patterns, making it less effective under rapid workload fluctuations. Finally, the design is restricted to single-region clusters and has yet to address distributed, multi-region deployment challenges, particularly in enabling WAN-aware KV cache sharing across geo-distributed sites.

As for future work, BanaServe can extend its capabilities along several promising directions. Hardware-aware scheduling and migration should be explored by incorporating device profiles that account for precision modes, interconnect bandwidth, and accelerator heterogeneity, enabling more efficient workload placement. Predictive orchestration, leveraging historical load traces and machine learning techniques such as reinforcement learning, could proactively adjust resource allocation and request routing to better handle workload spikes. Additionally, supporting multi-region scaling with WAN-optimized KV cache synchronization would allow BanaServe to achieve low-latency inference across geo-distributed deployments. Finally, integrating with container orchestration platforms like Kubernetes can facilitate realistic deployment, enabling the evaluation of scaling, scheduling, and placement policies in production environments.

\section*{Author Contributions}
All authors participated in extensive discussions and contributed to the conceptualization, analysis, and finalization of this manuscript. The specific contributions are as follows: Yiyuan He conceived the original research idea and implemented the experimental framework, conducted comprehensive experiments, and authored the primary content of the manuscript. Minxian Xu provided overarching guidance for the experimental methodology and manuscript development. Jingfeng Wu contributed to manuscript enhancement through detailed revisions, language refinement, and improvements in overall presentation quality. Jianmin Hu primarily assisted in the collection and verification of experimental data, ensuring the accuracy and reliability of results. Kejiang Ye and Chengzhong Xu provided critical insights into application scenarios and scalability considerations, and also contributed to the refinement of the manuscript. Authors from Alibaba Group AIOS Team (Chong Ma, Min Shen, Le Chen and Lin Qu) have provided Alibaba's large-scale experimental platform for performance validation and technical supports on developing BanaServe system. 

\section*{Acknowledgments}

\thanks{
This work is supported by National Natural Science Foundation of China under Grant 62572462,  Guangdong Basic and Applied Basic Research Foundation (No. 2024A1515010251, 2023B1515130002), Key Research and Development and Technology Transfer Program of Inner Mongolia Autonomous Region (2025YFHH0110) and Shenzhen Science and Technology Program under Grant JCYJ20240813155810014. We also thank Alibaba Group AIOS Team for providing large-scale platform and technical support.
}

\bibliography{banaserve}

\begin{thebibliography}{10}
\providecommand \doibase [0]{http://dx.doi.org/}%

\bibitem{achiam2023gpt}
Achiam J, Adler S, Agarwal S, et al. Gpt-4 technical report. {\it arXiv preprint arXiv:2303.08774.} 2023.

\bibitem{touvron2023llama}
Touvron H, Lavril T, Izacard G, et al. Llama: Open and efficient foundation language models. {\it arXiv preprint arXiv:2302.13971.} 2023.

\bibitem{claude}
Anthropic . Claude. \url{https://claude.ai};  2025.
\newblock Large language model.

\bibitem{monteiro2025nocodegpt}
Monteiro M, Branco BC, Silvestre S, Avelino G, Valente MT. NoCodeGPT: A No-Code Interface for Building Web Apps With Language Models. {\it Software: Practice and Experience.} 2025.

\bibitem{nguyen2025generative}
Nguyen-Duc A, Cabrero-Daniel B, Przybylek A, et al. Generative artificial intelligence for software engineering—A research agenda. {\it Software: Practice and Experience.} 2025.

\bibitem{zeng2025subkv}
Zeng Z, Zhang T, Lu Z, et al. Subkv: Quantizing Long Context KV Cache for Sub-Billion Parameter Language Models on Edge Devices. {\it Software: Practice and Experience.} 2025.

\bibitem{wen2025statuscale}
Wen L, Xu M, Gill SS, et al. StatuScale: Status-aware and elastic scaling strategy for microservice applications. {\it ACM Transactions on Autonomous and Adaptive Systems.} 2025\string;20(1)\string:1--25.

\bibitem{vaswani2017attention}
Vaswani A, Shazeer N, Parmar N, et al. Attention is all you need. {\it Advances in neural information processing systems.} 2017\string;30.

\bibitem{zhou2024survey}
Zhou Z, Ning X, Hong K, et al. A survey on efficient inference for large language models. {\it arXiv preprint arXiv:2404.14294.} 2024.

\bibitem{zhang2023h2o}
Zhang Z, Sheng Y, Zhou T, et al. H2o: Heavy-hitter oracle for efficient generative inference of large language models. {\it Advances in Neural Information Processing Systems.} 2023\string;36\string:34661--34710.

\bibitem{ge2023model}
Ge S, Zhang Y, Liu L, Zhang M, Han J, Gao J. Model tells you what to discard: Adaptive kv cache compression for llms. {\it arXiv preprint arXiv:2310.01801.} 2023.

\bibitem{zhong2024distserve}
Zhong Y, Liu S, Chen J, et al. DistServe: Disaggregating prefill and decoding for goodput-optimized large language model serving. In: 18th USENIX Symposium on Operating Systems Design and Implementation (OSDI 24). USENIX.  2024\string:193--210.

\bibitem{patel2024splitwise}
Patel P, Choukse E, Zhang C, et al. Splitwise: Efficient generative llm inference using phase splitting. In: 2024 ACM/IEEE 51st Annual International Symposium on Computer Architecture (ISCA). ACM.  2024\string:118--132.

\bibitem{sglang}
Zheng L, Yin L, Xie Z, et al. Sglang: Efficient execution of structured language model programs. In: . 37. NIPS.  2024\string:62557--62583.

\bibitem{vllm}
Kwon W, Li Z, Zhuang S, et al. Efficient memory management for large language model serving with pagedattention. In: Proceedings of the 29th symposium on operating systems principles. USENIX.  2023\string:611--626.

\bibitem{transformers}
Wolf T, Debut L, Sanh V, et al. Transformers: State-of-the-Art Natural Language Processing. In: Proceedings of the 2020 Conference on Empirical Methods in Natural Language Processing: System Demonstrations. ACL. Association for Computational Linguistics 2020; Online\string:38--45.

\bibitem{alpaca}
Taori R, Gulrajani I, Zhang T, et al. Stanford Alpaca: An Instruction-following LLaMA model. \url{https://github.com/tatsu-lab/stanford_alpaca};  2023.

\bibitem{longbench}
Bai Y, Lv X, Zhang J, et al. Longbench: A bilingual, multitask benchmark for long context understanding. {\it arXiv preprint arXiv:2308.14508.} 2023.

\bibitem{dao2022flashattention}
Dao T, Fu D, Ermon S, Rudra A, R{\'e} C. Flashattention: Fast and memory-efficient exact attention with io-awareness. {\it Advances in neural information processing systems.} 2022\string;35\string:16344--16359.

\bibitem{shah2024flashattention}
Shah J, Bikshandi G, Zhang Y, Thakkar V, Ramani P, Dao T. Flashattention-3: Fast and accurate attention with asynchrony and low-precision. {\it Advances in Neural Information Processing Systems.} 2024\string;37\string:68658--68685.

\bibitem{mukherjee2023orca}
Mukherjee S, Mitra A, Jawahar G, Agarwal S, Palangi H, Awadallah A. Orca: Progressive learning from complex explanation traces of gpt-4. {\it arXiv preprint arXiv:2306.02707.} 2023.

\bibitem{taming}
Agrawal A, Kedia N, Panwar A, et al. Taming Throughput-Latency tradeoff in LLM inference with Sarathi-Serve. In: 18th USENIX Symposium on Operating Systems Design and Implementation (OSDI 24). USENIX.  2024\string:117--134.

\bibitem{blitzscale}
Zhang D, Wang H, Liu Y, et al. BlitzScale: Fast and Live Large Model Autoscaling with O(1) Host Caching. In: 19th USENIX Symposium on Operating Systems Design and Implementation (OSDI 25). USENIX.  2025\string:275--293.

\bibitem{holmes2024deepspeed}
Holmes C, Tanaka M, Wyatt M, et al. Deepspeed-fastgen: High-throughput text generation for llms via mii and deepspeed-inference. {\it arXiv preprint arXiv:2401.08671.} 2024.

\bibitem{tensorrtllm}
{NVIDIA Corporation} . TensorRT-LLM: High-performance LLM inference library from NVIDIA. \url{https://developer.nvidia.com/tensorrt-llm}; .

\bibitem{nvidiadynamo}
{NVIDIA Corporation} . NVIDIA Dynamo: Distributed orchestration for large-scale LLM serving. \url{https://developer.nvidia.com/nvidia-dynamo}; .

\bibitem{wu2024loongserve}
Wu B, Liu S, Zhong Y, Sun P, Liu X, Jin X. Loongserve: Efficiently serving long-context large language models with elastic sequence parallelism. In: Proceedings of the ACM SIGOPS 30th Symposium on Operating Systems Principles. ACM.  2024\string:640--654.

\bibitem{hong2024flashdecoding++}
Hong K, Dai G, Xu J, et al. Flashdecoding++: Faster large language model inference with asynchronization, flat gemm optimization, and heuristics. {\it Proceedings of Machine Learning and Systems.} 2024\string;6\string:148--161.

\bibitem{hu2024tetriInfer}
Hu C, Huang H, Xu L, et al. Inference without interference: Disaggregate llm inference for mixed downstream workloads. {\it arXiv preprint arXiv:2401.11181.} 2024.

\bibitem{qin2024mooncake}
Qin R, Li Z, He W, et al. Mooncake: A kvcache-centric disaggregated architecture for llm serving. {\it arXiv preprint arXiv:2407.00079.} 2024.

\bibitem{hu2024memserve}
Hu C, Huang H, Hu J, et al. Memserve: Context caching for disaggregated llm serving with elastic memory pool. {\it arXiv preprint arXiv:2406.17565.} 2024.

\bibitem{chen2024kvdirect}
Chen S, Jiang R, Yu D, et al. KVDirect: Distributed Disaggregated LLM Inference. {\it arXiv preprint arXiv:2501.14743.} 2024.

\bibitem{feng2025windserve}
Feng J, Huang Y, Zhang R, Liang S, Yan M, Wu J. WindServe: Efficient Phase-Disaggregated LLM Serving with Stream-based Dynamic Scheduling. In: Proceedings of the 52nd Annual International Symposium on Computer Architecture. ACM. Association for Computing Machinery 2025; New York, NY, USA\string:1283–1295

\bibitem{hong2025semi}
Hong K, Chen L, Wang Z, et al. semi-PD: Towards Efficient LLM Serving via Phase-Wise Disaggregated Computation and Unified Storage. {\it arXiv preprint arXiv:2504.19867.} 2025.

\bibitem{Llumnix}
Sun B, Huang Z, Zhao H, et al. Llumnix: Dynamic Scheduling for Large Language Model Serving. In: 18th USENIX Symposium on Operating Systems Design and Implementation (OSDI 24). USENIX. USENIX Association 2024; Santa Clara, CA\string:173--191.

\bibitem{SpotServe}
Miao X, Shi C, Duan J, et al. SpotServe: Serving Generative Large Language Models on Preemptible Instances. In: ASPLOS '24. ACM. Association for Computing Machinery 2024; New York, NY, USA\string:1112–1127

\bibitem{uellm}
He Y, Xu M, Wu J, Zheng W, Ye K, Xu C. UELLM: A Unified and Efficient Approach for Large Language Model Inference Serving. In: Service-Oriented Computing: 22nd International Conference, ICSOC 2024, Tunis, Tunisia, December 3–6, 2024, Proceedings, Part I. Springer-Verlag. Springer-Verlag 2024; Berlin, Heidelberg\string:218–235

\end{thebibliography}

\end{document}